\documentclass[preprint,aps]{revtex4}
\usepackage{mathrsfs}
\pdfoutput=1%%\usepackage{collref}
\usepackage{graphicx}
\usepackage{multirow}
\usepackage{makecell}
\usepackage{booktabs}

\begin{document}

\title{New Developments in Laser-Based Photoemission Spectroscopy and its Scientific Applications: a Key Issues Review}

\author{Xingjiang Zhou$^{1,4,*}$, Shaolong He$^{2}$, Guodong Liu$^{1}$, Lin Zhao$^{1}$, Li Yu$^{1}$ and Wentao Zhang$^{3}$ }

\affiliation{
\\$^{1}$Beijing National Laboratory for Condensed Matter Physics, Institute of Physics, Chinese Academy of Sciences, Beijing 100190, China
\\$^{2}$Ningbo Institute of Materials Technology and Engineering, Chinese Academy of Sciences, Ningbo 315201, China
\\$^{3}$Department of Physics and Astronomy, Shanghai Jiao Tong University, Shanghai 200240, China.
\\$^{4}$Collaborative Innovation Center of Quantum Matter, Beijing 100871, China.
\\$^{*}$Corresponding author: XJZhou@iphy.ac.cn.
}

\date{May 30, 2017}

\begin{abstract}
The significant progress in angle-resolved photoemission spectroscopy (ARPES) in last three decades has elevated it from a traditional band mapping tool to a precise probe of many-body interactions and dynamics of quasiparticles in complex quantum systems. The recent developments of deep ultraviolet (DUV, including ultraviolet and vacuum ultraviolet) laser-based ARPES have further pushed this technique to a new level. In this paper, we review some latest developments in DUV laser-based photoemission systems, including the super-high energy and momentum resolution ARPES, the spin-resolved ARPES, the time-of-flight ARPES, and the time-resolved ARPES. We also highlight some scientific applications in the study of electronic structure in unconventional superconductors and topological materials using these state-of-the-art DUV laser-based ARPES. Finally we provide our perspectives on the future directions in the development of laser-based photoemission systems.
\end{abstract}

%\pacs{}% PACS, the Physics and Astronomy
\maketitle

\section{INTRODUCTION}
The physical properties of materials are primarily determined by their electronic structures, which are described by the three parameters of the electrons in a solid,  namely, energy (E), momentum (k)  and spin (s). Angle-resolved photoemission spectroscopy (ARPES), which probes the energy and momentum of electrons simultaneously, is a  direct tool in measuring the electronic structure of materials. The spin-resolved ARPES can further detect the spin states of materials. In the past three decades, the ARPES technique has been undergoing dramatic improvements in its performance, making it from a basic band structure mapping technique to a power tool in measuring the many-body effects and electron dynamics in the study of complex quantum materials.

\begin{figure}
\centering
\includegraphics[width=1.0\textwidth]{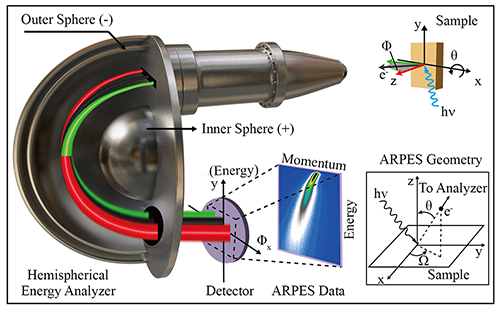}
\caption{\label{ARPES} Working principle of the angle-resolved photoemission with hemispherical electron energy analyzer. The inset shows the measurement geometry of the angle-resolved photoemission process. }
\label{fig:ARPES}
\end{figure}

The ARPES technique is based on the photoelectric effect, a quantum phenomenon discovered by Hertz in 1887\cite{Hertz1887} and theoretically explained by Einstein in 1905 by introducing the quantum nature of light\cite{Einstein1905}.  K. Siegbahn developed photoelectron spectroscopy to study the electronic structure in solids by employing intense X-ray sources and high-resolution electron energy analyzer, which won him the Nobel Prize in physics in 1981\cite{Siegbahn1957}.   The first demonstration  of band mapping by angle-resolved photoemission experiment was realized by Neville Smith et al. in early 1970's\cite{NSmith1973,NSmith1974}.  The underlying principle of ARPES is straightforward, as seen in the inset of Fig. \ref{fig:ARPES}.  The photoelectrons emitted by incident light move in a 2$\pi$ solid angle surrounding the sample. The electronic structure of the material  can be determined from the measured energy and the number of photoelectrons along different emission directions.

According to Einstein's quantum theory of photoelectric effect, electrons in solids are kicked out of materials by the incident photons, and the kinetic energy ($E_{kin}$) and momentum ($p$) of the photoelectrons are given by
\begin{equation}
E_{kin}=h\nu-W-E_{B}
\label{eq1}
\end{equation}
\begin{equation}
p=\hbar K=\sqrt{2mE_{kin}}
\label{eq2}
\end{equation}
in which $h\nu$ and W are the incident photon energy and the work function of solids, $E_{B}$ represents the binding energy, $K$ is the wave vector of photoelectrons in the vacuum, and m is the electron mass.  The parallel and perpendicular components of $p$ and $K$ are related to the emission angle of the photoelectrons ($\theta$, as defined in the inset of Fig. \ref{fig:ARPES}): $p_{\parallel}= \hbar K_{\parallel} = \sqrt{2mE_{kin}}\cdot \sin{\theta}$ and $p_{\bot} = \hbar K_{\bot} = \sqrt{2mE_{kin}}\cdot \cos{\theta}$.

The momentum of the electrons in Bloch state inside solids, $\hbar k$, is determined from the measured photoelectrons' momentum $p$. With a reasonable assumption that the parallel component of the momentum is conserved during photoemission process, the Bloch momentum parallel to the sample surface $k_{\parallel}$ is determined by $|k_{\parallel}+G|=K_{\parallel}=\sqrt{2mE_{kin}} \cdot sin\theta/\hbar$, in which $G$ is the corresponding reciprocal lattice vector. The momentum of the incident photons is neglected for low-energy photons. The perpendicular component of momentum $k_{\bot}$ cannot be determined directly since the momentum normal to the surface is not conserved because of the broken translation symmetry.  In order to determine the value of $k_{\bot}$, several  approaches have been proposed\cite{Hufner2003}. Since the bulk bands usually show obvious dispersion along $k_{\bot}$ direction,  $k_{\bot}$ can be determined from the periodic dispersions along this direction by photon-energy-dependent photoemission measurements.  One way is to compare experimental band dispersions with the calculated results at different $k_{\bot}$ and find out the proper value of k. The other way is based on the assumption of a free electron final state. In this case, the value of $k_{\bot}$ can be determined as:
\begin{equation}
k_{\bot}=\sqrt{\frac{2m}{\hbar^2}[(h\nu-W)cos^2\theta+V_0]}
\label{eq3}
\end{equation}
where m is the electron mass, W is the work function and V$_0$ is the so-called inner potential. Once the inner potential V$_0$ is determined from the photon-energy-dependent measurements, the value of $k_{\bot}$ can be determined. By measuring the energy and intensity of photo-emitted electrons at different emission angles,  ARPES is able to map the band structure, namely, determining both the energy and momentum of the electrons in solids.

Under sudden approximation, ARPES measures the single particle spectral function  that is the imaginary part of the Green's function $G(k,\omega)$\cite{Abrikosov1965,Hedin1969,Mahan1981,Rickayzen1991},
 \begin{equation}
A(k,\omega)=-\frac{1}{\pi}Im G(k,\omega)
\label{eq9}
\end{equation}
The Green's function is expressed as
\begin{equation}
G(k,\omega)=\frac{1}{\omega-\varepsilon_k-\Sigma(k,\omega)}
\label{eq6}
\end{equation}
Then the single particle spectral function can be written in terms of electron self-energy $\Sigma(k,\omega) = \Sigma'(k,\omega) +i\Sigma''(k,\omega)$ where $\Sigma'(k,\omega)$ and $\Sigma''(k,\omega)$ are the real and imaginary parts, respectively:
\begin{equation}
A(k,\omega)=-\frac{1}{\pi}\frac{\Sigma''(k,\omega)}{[\omega-\varepsilon_k-\Sigma'(k,\omega)]^2+[\Sigma''(k,\omega)]^2}
\label{eq7}
\end{equation}
In the photoemission process, $A(k,\omega)$ can be obtained by measuring the photoelectron intensity as a function of energy and momentum:
\begin{equation}
I(k,\omega)=I_0(k,\nu,A)f(\omega)A(k,\omega)
\label{eq8}
\end{equation}
where k is  momentum, $\omega$ is energy relative to the Fermi level, $I_0(k,\nu,A)$ is proportional to the matrix element $|M_{f,i}^k|^2$ which is related to the momentum of electrons (k), the energy ($\nu$) and polarization
of incident photons\cite{Hufner2003,Damascelli2003}; $f(\omega) = (e^{\omega/k_BT}+ 1)^{-1}$ is the Fermi-Dirac distribution function which indicates only occupied states are probed in the photoemission process.
By analyzing the measured $A(k,\omega)$ with the spectral function in Eq. \ref{eq7}, the real and imaginary parts of the electron self-energy can be extracted to study the many-body effects in solids, i.e., how electrons interact with other entities like electrons, phonons and etc. In the early time, because of limited energy and momentum resolutions, the application of ARPES is mainly limited to measure the gross band structure of materials. Since the late 1980s, the performance of ARPES has been dramatically improved thanks to the discovery of high temperature cuprate superconductors, followed by further exploration of the strongly correlated electron systems. ARPES has been developed into a leading experimental tool in condensed matter physics\cite{Damascelli2003,CampuzanoReview,XJZhou2007}.%%By probing the energy distribution of the photoelectron excited by light of photon energy $h\nu$, we can obtain the energy levels in solids according to the correspondence shown in Fig.\ref{fig:PESDOS}.

The principle of a modern ARPES is schematically shown in Fig. \ref{fig:ARPES}. The sketched hemispherical electron energy analyzer is equipped with a two-dimensional spatially-resolved detector and a multi-element electrostatic input lens. When it is operated in the angle-resolved mode, the photoelectrons within an angular span of $\sim$30 degrees along a window defined by the entrance slit (along x direction in Fig. \ref{fig:ARPES}) are collected by the angle-resolved lens and then spread to different positions along the x axis of a two-dimensional detector. Meanwhile, the photoelectrons are dispersed along the y axis of the detector as a function of their kinetic energy after passing through the hemispherical deflector. The combination of the angle-resolved lens and the two-dimensional detector makes it possible to measure multiple photoemission spectra simultaneously from different photo-emitting angles,  obtaining a two-dimensional snapshot of photoelectron intensity as a function of energy and momentum (Fig.  \ref{fig:ARPES})\cite{Beamson1990, Martensson1994}.  This multiple-angle collection of photoelectrons is a big jump of the ARPES technique from the previous one-angle-at-a-time measurement. In addition to the dramatic improvement of data acquisition efficiency,  the energy and momentum resolutions of the modern electron energy analyzers are also significantly improved; the typical energy resolution and angular resolution are better than 1 meV and 0.1 degree, respectively.

There are always strong demands in improving the performance of ARPES, including the energy and momentum resolutions, the data acquisition efficiency, and the stability of the system. The increasingly rich and deep physics in condensed matter physics pushes ARPES further to probe fine electronic structure under extreme conditions.  For example, it is still a challenge for ARPES to measure the superconducting gap of Sr$_2$RuO$_4$ with a superconducting transition temperature T$_c$$\sim$1 K\cite{Flichtenberg1994Sr2RuO4} and some heavy Fermion superconductors, taking CeCoIn$_5$ of a T$_c$ at 2.3 K as an example\cite{PMonthous2000CeCoIn5}.  The latest improvement of the ARPES has promoted it from a conventional band mapping tool to a cutting edge probe of the many-body effects in quantum materials\cite{XJZhou2007}. However, further improvements are urgently needed. Since the photon source plays a key role in dictating the ARPES performance,  great efforts have been made to develop better photon sources such as the synchrotron radiation light and monochromatic light from gas-discharge lamps.   The newly developed deep ultra-violet (DUV)(including ultra-violet(UV) and vacuum ultra-violet(VUV)) lasers are a novel and powerful source for the ARPES experiments\cite{Kiss2005,Koralek2006a,Koralek2007,GDLiu2008,Kiss2008}.   The DUV laser light sources have enabled ARPES with superior and unique advantages, pushing the photoemission technique to a new level and leading to tremendous high-quality and significant works in studying the fine electron structure of complex quantum materials such as unconventional superconductors and topological materials.

Generating a quasi-continuous-wave (CW) VUV laser light with a photon energy of $h\nu$ = 6.994 eV by the second harmonic process using KBe$_{2}$BO$_{3}$F$_{2}$ (KBBF) nonlinear crystal, VUV laser-based ARPES systems with an energy resolution better than 1 meV were developed\cite{GDLiu2008,Kiss2008}.  In the last decade, several other VUV laser-based photoemission systems have been developed. %. as shown in Fig. \ref{fig:IOPLaserARPES}.
In this brief review, we will cover the efforts in developing a series of VUV laser-based ARPES:  (1). VUV laser-based ARPES with super-high energy and momentum resolutions.  (2).  VUV laser-based spin- and angle-resolved photoemission spectroscopy(SARPES). (3).  VUV laser-based angle-resolved time-of-flight photoemission spectroscopy(TOF-ARPES). (4). VUV laser-based time-resolved ARPES (tr-ARPES).  We will highlight some scientific applications in the unconventional superconductors and topological materials by using these state-of-the-art VUV laser-based ARPES systems. Finally, we will provide some perspectives on future directions of the laser-based ARPES techniques.

%%To date, the best energy resolution achieved by using synchrotron radiation light or gas-discharge lamp is above 1meV.

\section{The Development of VUV Laser Light Source for Photoemission Spectroscopy}
\subsection{Basic Considerations of Laser Light as a Light Source}
Laser light source seems to be a natural choice for photoemission technique. Historically, laser light used as a photoemission light source can be traced back to very early time, mostly for time-resolved experiments\cite{Karlsson1996,Mathias2007},  including two-photon photoemission\cite{Haight1988,Nessler1998,Haight1994b}. The laser light was generated by using second-harmonic generation (SHG) in nonlinear optical crystal or high harmonic generation (HHG) technique in noble gas. Due to poor energy resolution,  low repetition rate,  low photon flux of the laser systems and the old-fashioned electron energy  analyzers,  these early laser-based photoemission techniques are  hard to provide information on the intrinsic electronic properties of the measured materials.  Until quite recently, near-ultraviolet (NUV) solid state source and HHG gas source became practical in the time-resolved pump$-$probe ARPES\cite{Perfetti2006,Schmitt2008,Rohwer2011,Petersen2011}. However, it is still impossible for these laser light sources to achieve super-high energy resolution on the order of a few meV.

%%for revealing the band dispersion and self-energy effects in solid\cite{Koralek2007,GDLiu2008} .

There are a number of basic requirements for a laser light to be used for a high resolution ARPES system:\\
(1). Proper photon energy: First, for the photoemission process to occur, the photon energy must be large enough to overcome the usual 4$\sim$5 eV work function of the material under measurement. The laser light must have a photon energy larger than 5 eV.  Second,  lower photon energy is desirable to achieve an enhanced bulk sensitivity,  high photoemission cross section, reduced extrinsic background,  high momentum resolution and a better selectivity in perpendicular wavevector k$_{\bot}$. For the photon energy between 20 and 100 eV frequently used in ARPES measurements, the mean free path of the photoelectrons is typically$ <$ 10${\AA}$  \cite{Seah1979}, which renders the technique rather surface sensitive. For the lower photon energy between 6 and 15 eV, the corresponding electron mean free path is enhanced to $\sim$ 20$\sim$100{\AA}depending on the material\cite{Koralek2007,GDLiu2008,Seah1979}.
Third, one should consider the momentum space that can be covered at a given laser photon energy; lower photon energy will lead to a limited momentum space. Taking all the above factors into consideration, a possible proper range of photon energy for a super-high resolution ARPES can be 6$\sim$15 eV.\\
(2). High photon flux: The laser light must have high enough intensity in order to get decent photoemission signal to achieve sufficiently high signal-to-noise ratio and high data acquisition efficiency;\\
(3). Narrow bandwidth: The bandwidth of the laser light will be a major factor in determining the overall instrumental energy resolution. To get high energy resolution, a narrow bandwidth of the laser light is necessary. \\
(4). Continuous-wave (CW) or quasi continuous-wave laser light source:  Photoemission process involves the space charge effect that may shift the energy position, broaden the linewidth and distort the lineshape of the photoemission spectra\cite{XJZhou2005,Passlack2006,Graf2010}.  This effect becomes prominent when there is a large number of photons in one pulse, such as in the pulsed laser.  It is preferable to use CW or quasi-CW laser light ($\sim$100 MHz) in order to minimize the space charge effect to get high energy resolution and intrinsic signal.\\
(5). Compatible electron energy analyzer: The relatively low kinetic energy of photoelectrons when using laser light puts a stringent demand on the electron energy analyzer. First, it requires the angular mode of the electron energy analyzer to work properly at such a low electron kinetic energy. Second, it should have a super-high energy resolution in order to take full advantage of the narrow bandwidth of the laser light\cite{GDLiu2008,Kiss2008}.\\
(6). Variable polarizations:  Controllable polarizations of laser light can be used to take advantage of  the photoemission matrix element effect and disentangle the orbital characters of the energy bands\cite{Hufner2003,Damascelli2003}.  The circular polarizations can be used to study magnetism or test time-reversal symmetry breaking.\\
(7). Variable ($<$1 mm) beam spot size: Small spot size is desirable for improving the performance of the electron energy analyzer and measuring on a small area of samples. On the other hand, small spot size tends to give rise to strong space charge effect.  In order to get a good balance and depending on different materials measured, it is helpful to be able to tune the laser light spot size.\\
(8). Long term stability:  Laser light source should be stable during the ARPES measurements that usually last for a couple of days.\\
(9). Compact and simple construction: This will lead to a considerable reduction in building and operating costs as well as lab space requirements.\\

For the time-resolved ARPES (tr-ARPES), in addition to the requirements listed above, the pulse duration, the repetition rate, and the wavelength tunability of the laser light should also be considered\cite{Orenstein2012}.

\subsection{Generation of Laser Light for Photoemission}
In general, the UV and VUV laser light required for the photoemission process can be generated through various frequency up-conversion techniques that are primarily divided into two categories. One utilizes the second order nonlinear processes (second harmonic generation, SHG) in the non-linear optical crystals (NLOC) to produce deep-UV laser light. These laser light sources are mainly used in the high-resolution ARPES system\cite{Koralek2007,GDLiu2008,Kiss2008,Tamai2013,Carpene2009,Schmitt2011,Faure2012,Sobota2012}. The other is based on the higher order nonlinear processes (high harmonic generation, HHG) in the noble gases to generate extreme ultra-violet (EUV) and even X-ray ultra-violet (XUV) laser light. Such a short wavelength radiation is mostly used in the tr-ARPES apparatus\cite{Rohde2016,Frietsch2013,Rohwer2011,Dakovski2010,Mathias2007,Siffalovic2001}. Usually  the second-order nonlinear process has a higher generation efficiency than the HHG in noble gas. Recently, free electron laser (FEL) has emerged as a versatile light sources for the tr-ARPES as the probe beam. Table \ref{fig:Laserlist} lists some typical types of laser light source used for ARPES systems.

\begin{figure}
\begin{center}
\includegraphics[width=0.6\columnwidth,angle=0]{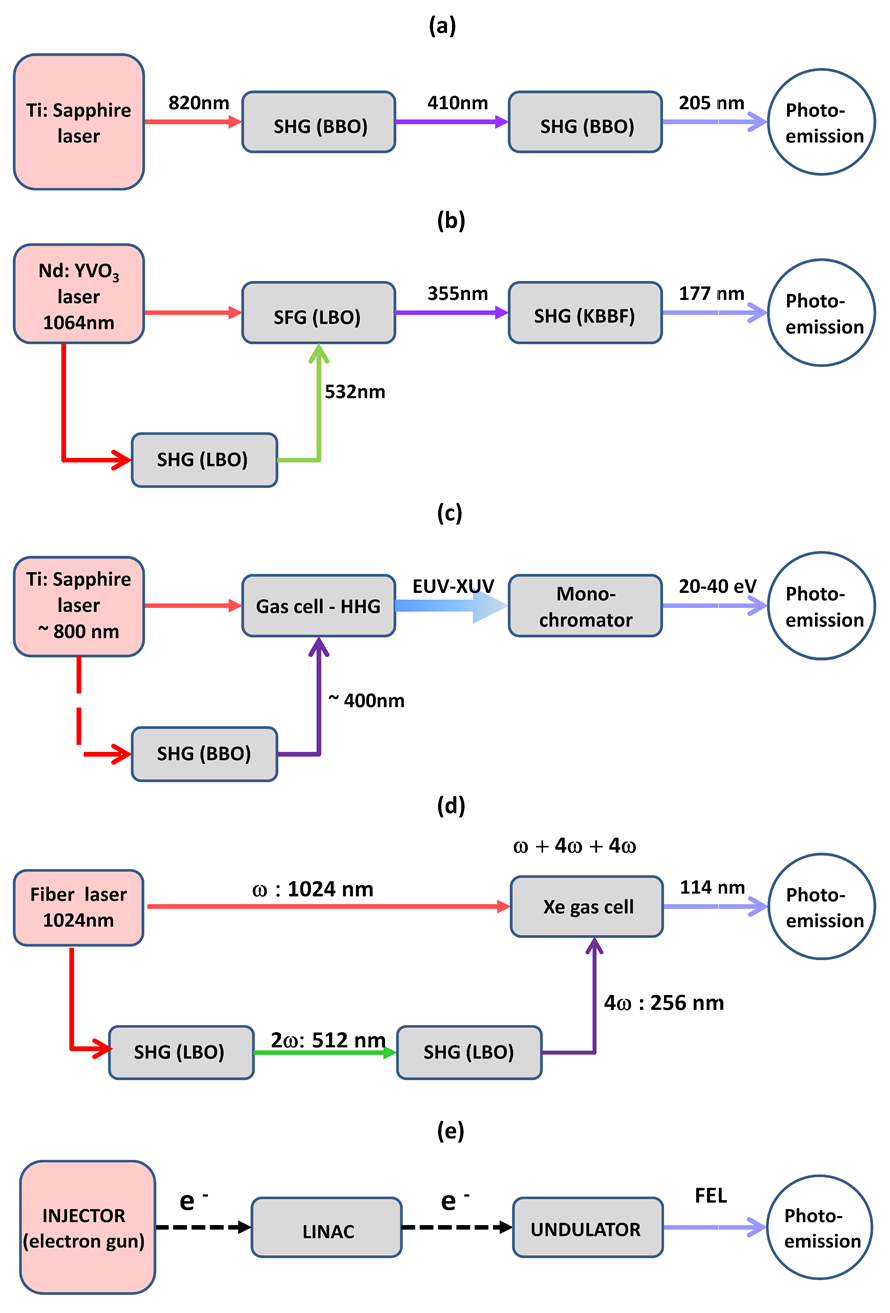}
\end{center}
\caption{ Schematic layout of the working principles for different laser sources. (a).  For pulsed ~100 kHz UV laser (fs-version and ps-version ) and CW UV laser. (b).  For Quasi-CW VUV laser $\sim$100 MHz and pulsed $\sim$100kHz. (c).  For HHG laser: 1$\sim$100kHz. (d).  For the Resonance-type HHG: $\sim$200 kHZ - 50MHz. (e).  For free electron laser.}
\label{fig:Laser}
\end{figure}

%%\cite{Wikipedia}
Both SHG in crystals and HHG in noble gases originate physically from the nonlinear optical process (NLO) in which high intensity laser light interacts with a nonlinear media, that is,  the dielectric polarization P responds nonlinearly to the electric field E of the light.
%%Equation
\begin{equation}
P(t)=\varepsilon_{0}(\chi^{(1)}E(t)+\chi^{(2)}E^{2}(t)+\chi^{(3)}E^{3}(t)+...)
\label{eq3}
\end{equation}
where the coefficients $\chi^{(n)}$ are the n$^{th}$ order susceptibilities of the medium.  In general, $\chi^{(n)}$ is an (n + 1)$^{th}-$rank tensor. As a noble gas atom is centrally symmetric, %%Therefore, for a gas molecular being central symmetric,
the even order susceptibilities become zero.  For HHG in gas, only odd multiple of the fundamental frequency can be produced. While for the non-central symmetric crystals,  $\chi^{(2)}$ does not equal to zero, so SHG can take place. Fig. \ref{fig:Laser}a-e schematically shows the working principles of different laser sources that are utilized in ARPES laboratories all over the world\cite{Koralek2007,GDLiu2008,Kiss2008,Rohde2016,Frietsch2013,Rohwer2011,Dakovski2010,Mathias2007,Siffalovic2001,Carpene2009,Faure2012,Schmitt2011,Sobota2012,Kirchmann2008,ZJXie2014a,CLWang2016,Vishik2010,Hellmann2010,Allaria2012}.

\subsubsection{SHG: Solid State Deep UV Laser light source}
The most popular and reliable approach is to employ the SHG technique to realize a deep-UV laser light source for ARPES application.  This technique is easy to implement because the laser light beams do not need to be separated and overlapped spatially and temporally \cite{Koralek2007,GDLiu2008,Kiss2008,Carpene2009,Faure2012,Schmitt2011,Sobota2012,Kirchmann2008,ZJXie2014a,CLWang2016,Vishik2010}. The most important (key) element is the NLO in the frequency up-conversion of the solid state laser light source.  Specifically, SHG can be described by the following equation.
%%Add Equation
\begin{equation}
P=\alpha E+\beta E^{2}=\frac{1}{2}\beta E^{2}_{0}+\alpha E_{0}cos(\omega t)+\frac{1}{2}\beta E^{2}_{0}cos(2\omega t)
\label{eq4}
\end{equation}
Under the non-depleted pump approximation\cite{XZhang2012}, the theoretical VUV output power can be calculated by the following formula.
%%Equation
\begin{equation}
P_{2\omega}= \frac{8\pi^{2}d^{2}_{eff}L^{2}I_{\omega}}{\varepsilon_{0}n^{2}_{\omega}n_{2\omega}c\lambda^{2}_{\omega}}P_{\omega}
\label{eq5}
\end{equation}
where $d_{eff}$ is the effective NLO coefficient of an NLO crystal; L is the optical path in the crystal; $I_{\omega}$ is the peak power density of the fundamental beam in NLO crystal; $\varepsilon_{0}$ is the vacuum permittivity, $n_{\omega}$ and $n_{2\omega}$ are the refractive indices of NLO crystal at frequency $\omega$ and $2\omega$; c is the light speed in vacuum, ${\lambda}_{\omega}$ is the fundamental wavelength, and $P_{\omega}$ and $P_{2\omega}$ are the powers of $\omega$ and 2$\omega$ beams, respectively.

\begin{figure}
\begin{center}
\includegraphics[width=1.0\columnwidth,angle=0]{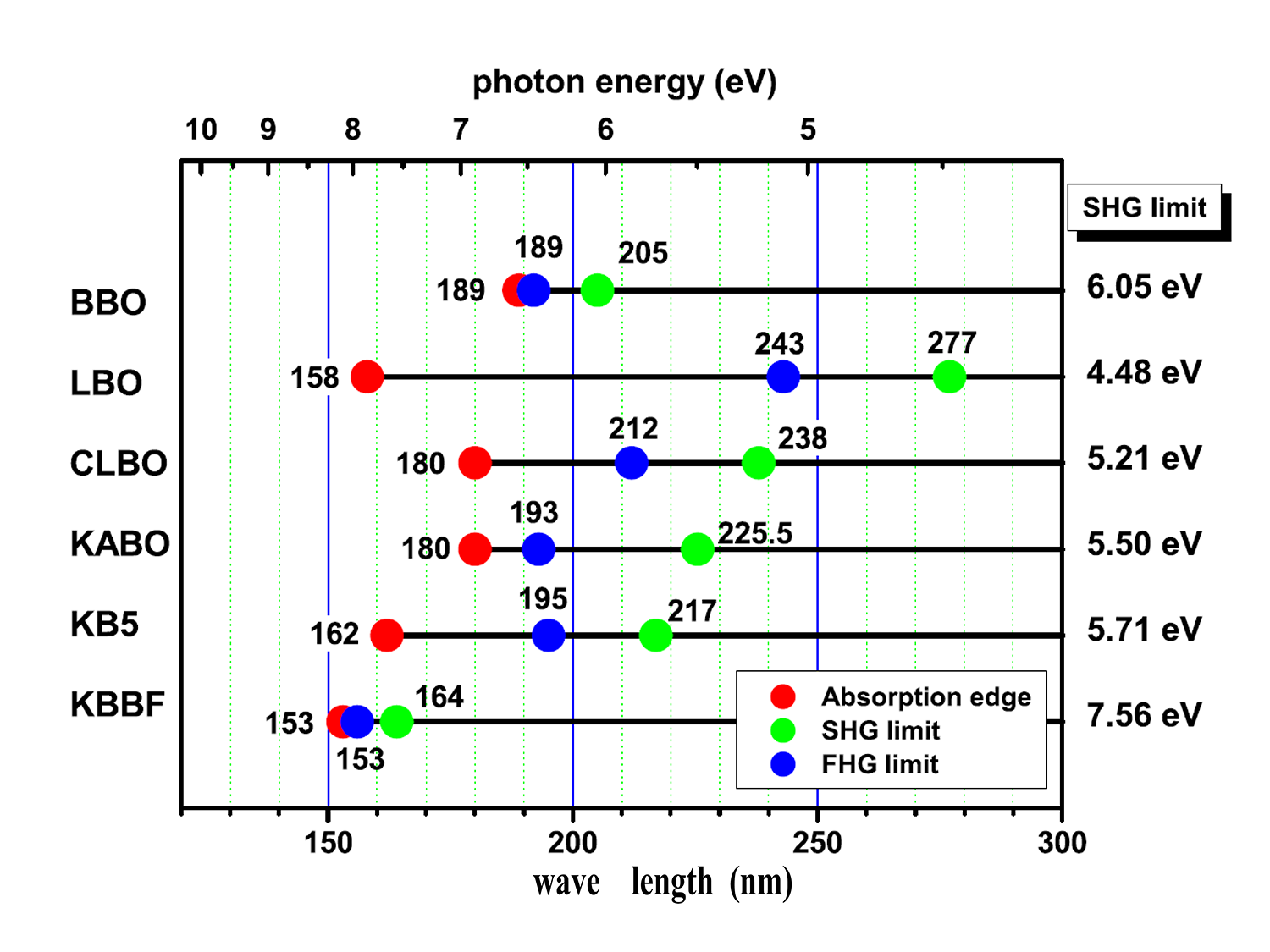}
\end{center}
\caption{A summary of some typical NLO crystals and their SHG limit, FHG (fourth harmonic generation) limit, and absorption edge\cite{GDLiu2008}. The shortest SHG wavelength (highest SHG energy) is only available for KBBF crystal among all the NLO crystals.}
\label{fig:NLOC}
\end{figure}

A VUV NLO crystal that generates laser light applicable for photoemission should have a high NLO coefficient, a short absorption edge below 200 nm, and a moderate birefringence (0.07$\sim$0.10).   Fig.  \ref{fig:NLOC} summarizes the shortest second-harmonic generation SHG wavelength available for the typical and well-developed NLO crystals\cite{GDLiu2008}.  Although BaB$_{2}$O$_{4}$ (BBO) and LiB$_{3}$O$_{5}$ (LBO) are usually used for SHG from visible to ultraviolet all the way to 200 nm, their performance below 200 nm is rather poor. In the case of BBO, since the absorption edge is at 189 nm, it can only achieve SHG output not shorter than 210 nm.  For LBO, although the transparent range is wide, the phase matching is limited by its small birefringence.  As a result, the shortest SHG output wavelength of LBO is only 276 nm under phase-matching conditions.

Figure \ref{fig:Laser}a schematically shows a $\sim$6 eV laser light generation that has become a popular and robust solution of light source in many lab-based tr-ARPES and high resolution ARPES systems\cite{Koralek2007,Tamai2013,Carpene2009,Schmitt2011,Faure2012,Sobota2012}. This approach commonly utilizes a cascade of two frequency doubling stages in nonlinear BBO crystals through type I phase matching. The first SHG laser light is used as the input of the second harmonic stage. Finally, the ~$\sim$200 nm UV laser light can be achieved through the fourth harmonic generation (2$\omega$+2$\omega$: FHG) out of a $\sim$800nm Ti:sapphire laser light with a high repetition rate of hundreds of kHz or MHz. Basically, a lower-limit wavelength 205 nm is restricted by the phase matching condition in this scheme.

\begin{figure}
\begin{center}
\includegraphics[width=1.0\columnwidth,angle=0]{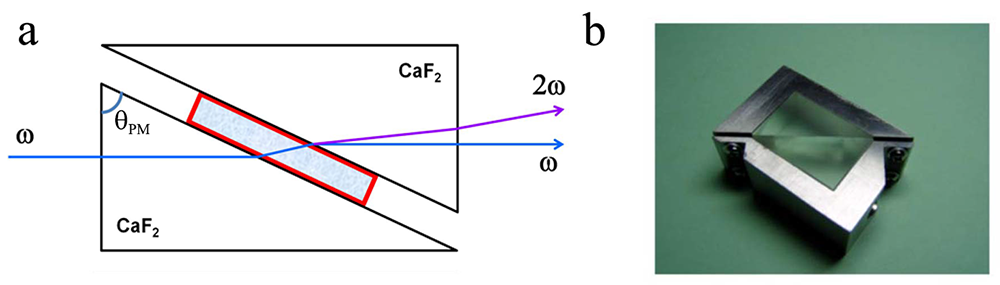}
\end{center}
\caption{(a). The schematic sandwich structure of an optically contacted, prism-coupled KBBF crystal device.  (b).  A photo of a typical KBBF-CaF$_2$ prism-coupled device.}
\label{fig:KBBF}
\end{figure}

KBBF is a novel UV NLO crystal that can break the 200 nm wavelength barrier and enter the VUV region. It has a large NLO coefficient of d$_{11}$ = 0.49 pm/V, good birefringence of n=0.072, and a very short transmission cutoff wavelength of 152 nm. So far, it is the only crystal available that enables phase matching below 160 nm for SHG. However, a major obstacle of using KBBF lies in its platelike nature. It is difficult to grow large-sized KBBF crystal that is suited to be cut at the phase-matching angle. To solve this problem, Chen et al. developed a prism-coupled technique (PCT) device, as schematically shown in Fig. \ref{fig:KBBF}\cite{CTChen2002}, which is composed of a KBBF crystal sandwiched in between two CaF$_{2}$ prisms, or  one SiO$_{2}$ and one CaF$_{2}$ prism,  with a proper apex angle. By using this KBBF-PCT device,  laser light with 172.5 and 163.3 nm wavelengths have been demonstrated which are the shortest achieved so far by nonlinear crystals for SHG and sum-frequency mixing (SFM), respectively.

Recently, the continuous-wave laser light sources have also been developed and used in ARPES experiments as reported for a 6.05 eV (205 nm) laser light source\cite{Tamai2013}.  The CW laser light used in ARPES has two additional advantages: one is the extremely narrow bandwidth of the light due to its infinite pulse duration;  the other is theoretically zero space charge effect. CW laser light sources based on KBBF with a power level of 1.3 mW at 191 nm (6.49 eV)\cite{Scholz2012} and  15 mW at 193 nm (6.43 eV)\cite{Scholz2013} have also been reported. The $\sim$190 nm laser light source can cover larger momentum space than that by 205 nm (6.05 eV) laser light in ARPES measurements.

\subsubsection{High Harmonic Generation}
VUV or EUV laser light sources based on HHG process have been widely used in the tr-ARPES experiments, producing a lot of significant results in unveiling the non-equilibrium dynamics of electrons in charge-density-wave (CDW) materials,  magnetic surface, and ultrafast atom-specific electronic dynamics\cite{Rohwer2011,Frietsch2013,Ishizaka2011}.  The HHG process occurs in a gas cell or gas jet when the noble gas atoms are illuminated by a near-IR laser beam or near-UV laser beam and sometimes their combination with the intensity reaching up to 10$^{14}$ W$\cdot$cm$^{-2}$ in the infrared regime.  The earliest observation of this phenomenon is back to 1990s\cite{Corkum1993}. The basic mechanism can be qualitatively described by a 3-step process\cite{Frietsch2013,Bauer2005}.  In the first step, the bound electron becomes free after being ionized by the high electric field of the laser light through multi-photon ionization or tunnel ionization processes. Second, the electron is accelerated by the laser light field and then returns to the vicinity of its parent ion when the laser light field reverses. Third, the electron may recombine with the ion, emitting an high energy photon. Its energy is an odd multiple of the IR photon energy. The recombination usually has a very low probability on the order of 10$^{-6}$. It leads to an extremely low frequency up-conversion efficiency from IR to XUV laser light.

%%, as shown in Fig. \ref{fig:3Step}

In the HHG process, the maximum achievable photon energy is limited only by the kinetic energy that the electron can accumulate during the optical half cycle\cite{Bauer2005}. Therefore, such a process is most effective with few-cycle femtosecond pulses operating at a lower repetition rate of about 1-100 kHz. Since HHG is a highly nonlinear process, the efficiency and the cut-off energy depend sensitively on a number of experimental factors\cite{Frietsch2013,Bauer2005}, including gas species and pressure, laser light properties (wavelength, the pulse width, power and repetition rate), and the focusing conditions (beam profile, the focusing geometry and the interaction length).

It is worth mentioning that a laser light source with a photon energy of 11 eV was recently developed  which combines the merits of narrow bandwidth in solid state frequency doubling and accessibility of short wavelength in gas HHG in some sense\cite{Hellmann2012}.  As shown in Fig. \ref{fig:Laser}d, it employs a resonant-type HHG process in Xenon which reduces the pumping peak power from MW usually used down to KW level. This special VUV laser light source can run  at tunable repetition rate between 100 kHz and 10 MHz, provide a photon flux around 2$\times 10^{12}$ photons/s, and enable ARPES with energy and momentum resolutions better than 2 meV and 0.012 ${\AA}^{-1}$, respectively.  In particular, with this laser light, the first Brillouin zone of most high temperature superconductors can be accessed, since the photoemitted electron momentum can reach up to 1.2 ${\AA}^{-1}$.

\subsubsection{Free-Electron Lasers}

Free-electron lasers (FELs) have been recognized as a promising source for the generation of highly brilliant and tunable EUV/X-ray laser radiation, which can be used in the time-resolved photoemission (tr-PES) experiments\cite{EASeddon2017}. FEL does not require a gain medium of gas, liquid or solid like in general lasers. It directly converts the kinetic energy of the  relativistic high energy ''free" electrons (drawing out of accelerator) into the coherent radiation after such  electrons pass through an undulator. In FELs, the electrons are forced to emit coherently by prolonged
interaction with the electromagnetic field emitted within the undulator. Depending on the origin of the  electromagnetic wave initiating the bunching process, there are two types of FEL operation mechanism: self-amplified spontaneous emission (SASE)\cite{Hellmann2010,Hellmann2012} and laser seeded mechanism (e.g. high-gain harmonic generation  (HGHG), Echo-Enabled Harmonic Generation (EEHG))\cite{Allaria2012}. In the SASE mechanism like FLASH at DESY, Germany\cite{http1},  FEL usually shows relatively large shot-to-shot intensity and photon-energy fluctuations and the limited  longitudinal coherence. Instead, the laser seeded mechanism like FERMI FEL-1 at Elettra ST, Italy, has the  shot-to-shot wavelength stability, low-intensity fluctuations, close to transform-limited bandwidth,  transverse and longitudinal coherence and full control of polarization\cite{Allaria2012,Allaria2012b}. Besides FLASH and FERMI, there  are another two currently operational X-ray FEL facilities: LCLS\cite{http7} at SLAC in California,USA, and SACLA\cite{http8}  at RIKEN in Japan. In addition, some new X-ray FEL facilities are under contruction, including  the European XFEL\cite{http9}at DESY in Germany, the SwissFEL\cite{http10} at the Paul Scherrer Institute (PSI) in Switzerland,  the PALX-FEL at Pohang Accelerator Laboratory\cite{http11} in Korea and the SXFEL\cite{Wang2016} at the Shanghai Institute  of Applied Physics (SINAP) in China.

To see how a FEL works, in Fig. 7, we show the  schematic layout of the first XUV and soft X-ray FEL  apparatus FLASH\cite{Hellmann2010,Hellmann2012,http1}. At the first step, electron bunch patterns are generated by a radio-frequency gun according to the user-preferred operation mode. The electron bunches are compressed longitudinally and accelerated to 1.25 GeV in the linear accelerating structures. Finally they are guided into a long undulator (30 meters) to generate the FEL radiation via the SASE process. At present, FLASH can routinely produce very bright and ultrashort-pulse coherent radiation with a photon energy between 26 and 300 eV, a pulse width of 30$-$150 fs, a master repetition rate up to 10 Hz (250 kHz single-pulse repetition rate in the macrobunch mode)  and a pulse energy up to 130 $\mu$J.

\subsection{Evaluation of Laser light as a Source for Photoemission }
For ARPES, there are basically three types of light sources which have been successfully and routinely used, including rare gas discharge lamps, synchrotron radiation and laser light source. In comparison with discharge lamp and synchrotron radiation, laser light source has many advantages but also some limitations. At current stage, these three light sources are complementary for photoemission experiments.

The earliest photoemission light sources are the noble gas discharge lamps (helium, neon and xenon) with a few discrete photon energies available of 21.2 eV, 40.8 eV, 16.85 eV and 8.5 eV\cite{VUV5k,UVLS,Souma2007}. These lamps are compact, practical and particularly affordable to most ARPES labs. A typical example is the modern Helium lamp employing the microwave plasma technique\cite{VUV5k}. Such a Helium lamp is still commonly used in many ARPES labs  since it can provide rather high flux of 21.2 eV photons to meet the ordinary requirements for ARPES measurements. It is free from space charge effect due to its CW emission nature. However, it has a few obvious shortcomings: insufficient photon flux for high resolution measurements,  fixed photon energy,  fixed or no polarization,  relatively large beam spot size, and leakage of helium gas into the ARPES measurement chamber.

In the last three decades, the continuous advancement of synchrotron technology, especially the emergence of the third generation synchrotron light sources,  and the technical progress in two-dimensional detection-type hemispherical electron energy analyzers\cite{DA30,Phoibos,MBSAnalyzer} have dramatically improved the performance of ARPES,  making it a leading technique for probing electronic structure and many-body interactions in condensed matter physics. The third generation synchrotron sources are featured by very brilliant beam and its continuous tuning in a large photon energy range from VUV to soft-X-ray regions. The beam can be focused to sub-millimeter down to sub-micrometer in size, thus improving the performance of the electron energy analyzer and facilitating ARPES measurements on small or inhomogeneous samples.  It has tunable polarizations (linear and circular) that helps to take advantage of the matrix element effects and identify orbital character of the energy bands. Despite the outstanding capabilities of synchrotron-based ARPES, it remains a challenge to achieve a super-high energy resolution better than 1 meV because for synchrotron light source, the narrow bandwidth is obtained at a cost of photon flux. When the bandwidth reaches 1 meV level, the photon flux usually becomes  too low to perform practical ARPES measurements.

The newly developed DUV lasers provide an efficient, compact and low-cost light source for the laboratory-based ARPES systems with many unique advantages\cite{Koralek2006a,GDLiu2008,Kiss2008,Vishik2010}. The extremely high and stable photon flux greatly enhances the data acquisition efficiency and data quality.   The narrow bandwidth makes it possible to achieve super-high energy resolution in ARPES measurements. In particular,  different from synchrotron radiation source, the bandwidth and photon flux of the laser light source  are not correlated and can be independently optimized. The utilization of VUV laser light (6.994 eV) has successfully broken the barrier of 1 meV super-high energy resolution that has been a long-sought dream in ARPES technique\cite{GDLiu2008,Kiss2008}.
In addition, the relatively low photon energy gives rise to high momentum resolution and enhanced bulk sensitivity. It is also easy to control the polarization of the laser light and tune the beam spot size.
 %The superior performance of laser light has definitely elevated the photoemission technique into a new level.  In some important figures of merit like the energy resolution, photon flux and bulk sensitivity, a narrow-band laser light source surpasses conventional ARPES sources like synchrotron radiation or discharge lamps by at least one order of magnitude\cite{GDLiu2008}.

Although typical HHG laser light is still hard to achieve high energy resolution in  ARPES experiments,  recent advances in modified HHG techniques have resulted in laser light at photon energies of 10.5 eV\cite{Berntsen2011} and 11 eV\cite{Berntsen2011,YHe2016}.  They have narrow band-width and high photon flux that are comparable to those in solid state UV/VUV laser light.  Moreover, for time-resolved photoemission, the laser light with femtosecond-scale pulse width are viable light sources\cite{Frietsch2013,Rohde2016,Faure2012,Perfetti2006,Schmitt2008,Rohwer2011,Smallwood2012a}.

%%Currently only laser source can provide both the necessary time and energy resolution to map non-equilibrium states of the band structure.

\section{Development of VUV Laser-Based ARPES Systems}
\subsection{VUV Laser-Based ARPES with Super-High Energy and Momentum Resolutions}
Figure \ref{fig:LaserARPES} shows a typical VUV-laser based ARPES system. The successful development stems from two major advancements: generation of 6.994 eV VUV laser light and advent of new electron energy analyzer that its angular mode can work properly at a low electron kinetic energy.  The new VUV laser-based ARPES system exhibits superior performance, including super-high energy resolution better than 1 meV,  high momentum resolution, super-high photon flux, and much enhanced bulk sensitivity\cite{GDLiu2008}.

A central part of the system is the picosecond quasi-CW VUV laser light source with a photon energy of 6.994 eV and a bandwidth of $\sim$0.26 meV.   It is achieved from the SHG using a novel nonlinear optical crystal KBBF (see Fig. \ref{fig:Laser7eV})\cite{CTChen1996,Togashi2003,CTChen2009}.  A 355 nm commercial frequency-tripled Nd:YVO$_4$ laser (Vanguard, Spectra Physics) is introduced into the frequency doubling vacuum chamber and focused onto a KBBF crystal by a pair of reflection mirrors and a quartz lens. Here the key element is the optically-contacted KBBF-CaF$_{2}$ prism-coupled device where the SHG occurs. The generated 177.3 nm VUV laser light is reflected by two high reflective mirrors and focused by an anti-reflection coated CaF$_{2}$ lens onto the sample position in the ARPES analysis chamber. Behind the second reflection mirror, a half-wave plate and a quarter-wave plate working at 177.3 nm are placed for getting  linear and circular polarizations of the light. The analysis chamber is isolated in vacuum from the optical chamber by a CaF$_{2}$ window; this avoids using complicated differential pumping device to keep the ultra-high vacuum in the analysis chamber.

Compared with 6 eV UV laser light developed for ARPES before\cite{Koralek2007}, the advantage of the 6.994 eV VUV laser light lies in both the superior energy resolution and relatively large momentum space coverage. The linewidth of the VUV laser light source is $\sim$0.26 meV which is significantly narrower than $\sim$5 meV for the UV laser light\cite{Koralek2007}. Combined with the resolution of the electron energy analyzer, a high resolution of 0.36 meV have been achieved for the VUV laser-based ARPES\cite{GDLiu2008,Kiss2008}.  At the photon energy of h$\nu$ =6.994 eV, the maximum in-plane momentum covered is 0.84 ${\AA}^{-1}$ taking a work function of 4.3 eV while it is 0.67 ${\AA}^{-1}$ for h$\nu$ =6 eV. The advantage of larger photon energy is obvious here because 6.994 eV laser light can nearly reach the important ($\pi$,0) region of high-T$_c$ cuprate superconductors, the iron-based superconductors and many other transition metal oxides.

A natural question that arises when interpreting the laser-based ARPES results is the validity of the so-called sudden approximation, which assumes that the photoelectron leaves the sample before the relaxation of the created photo-hole and neglects the interaction between the outgoing photoelectron and the remaining system\cite{Hedin1998,Hedin2002,JDLee1999}.  For a high photon energy, this is indeed a good approximation as demonstrated in high temperature cuprate superconductors\cite{Randeria1995}. When the kinetic energy of photoelectron is lower than 3 eV, a typical value for laser-based ARPES, whether the sudden approximation still holds remains to be examined. Koralek et al.\cite{Koralek2006a} directly compared the band dispersions taken at a low photon energy (6 eV) with those at high photon energy and found that the main features are identical. Their results support the validity of the sudden approximation for laser-based ARPES measurements.

While the VUV laser-based ARPES\cite{GDLiu2008,Koralek2007,Kiss2008} possesses a host of advantages over synchrotron radiation source in terms of energy resolution, momentum resolution, photon flux and bulk sensitivity, it has an obvious drawback in its single photon energy. Due to the photoemission matrix element effects,  a single fixed photon energy may make ARPES measurements miss some energy bands and, in the worst case, may not be able to work on some materials. Moreover, for some materials with significant k$_z$ dispersion, different photon energies are necessary to get access to electronic structure at different k$_z$s. Therefore, the upgrade of laser light source from single photon energy to tunable photon energy is highly necessary in high resolution ARPES community.

The availability of the new NLO crystal,  KBBF,  provides an opportunity to construct a tunable solid state laser light source for the ARPES technique.  Fig. \ref{fig:Tunable} shows the schematic layout of a tunable laser system with a photon energy ranging from 5.9 to 7.1 eV\cite{SJZhang2014,YZhang2017}. The tunable Ti:Sapphire laser light (Tsunami ps-version 80 MHz, Spectra Physics) is pumped by a 15 W CW Millennia laser at  532 nm. Its wavelength can be tuned almost continuously from 700 to 850 nm. The pulse duration is 2 ps at 800 nm. The tunable IR beam is first focused inside the BBO-SHG device that is composed of two head-to-head arranged BBO crystals to enhance the SHG output and keep the beam direction during wavelength change. The SHG UV beam is introduced into the frequency doubling vacuum chamber and focused onto KBBF crystal through a quartz window by a reflection mirror and a quartz lens. Then the FHG beam is generated inside the key KBBF-PCT element via a SHG process. The achieved 175 - 212 nm DUV laser light is reflected by two high reflective mirrors and focused by an anti-reflection coated CaF$_{2}$ lens onto the sample position in the ARPES analysis chamber. Between the analysis chamber and optical chamber, a CaF$_{2}$ window is set to get a vacuum isolation.  We note that a similar ARPES system using tunable laser light source was also reported by Kaminski's group\cite{RJiang2014}. They used a hydrothermally-grown KBBF crystal to realize a tuning of photon energy between 5.3 and 7 eV with a repetition rate of 76 MHz.  Although these tunable ARPES can tune the photon energy over a limited energy range, they have proven to be very valuable in ARPES measurements\cite{YZhang2017,LHuang2016,YWu2016}.

\subsection{VUV Laser-Based Spin-Resolved ARPES}
In order to determine complete electronic states  of materials, one needs to have information about all three parameters of electrons, namely, energy, momentum and spin.  While ARPES can measure the energy and momentum of electrons,  it does not have the  spin information. The spin state in materials has become more and more important due to rapid development in both fundamental studies and potential applications, especially for topological quantum state\cite{Hasan2010, SCZhang2011} and spintronics\cite{Wolf2001}.   Therefore, a great deal of efforts have been devoted to  the development of the spin- and angle-resolved photoemission spectroscopy (SARPES), which combines ARPES with the spin detectors.

Many types of spin detectors have been developed for SARPES by using Mott scattering\cite{Mott1929,Mott1932}, polarized low-energy electron diffraction (PLEED) from single crystals\cite{Kirschner1979}, diffuse scattering\cite{Unguris1986}, and very low energy electron diffraction (VLEED) mechanisms\cite{Tillmann1989}. Among these spin detection schemes, the VLEED spin detector, developed in 1989 for the first time\cite{Tillmann1989}, has the highest efficiency due to its higher scattering probability of the low energy electrons ($E_{k}\sim$ 6-10 eV) compared with that of Mott scattering ($E_{k}\sim$ 20-100 keV).  After solving the problem of quick degradation of the ferromagnet target\cite{Bertacco1998}, the VLEED spin detector has recently been put into practical use  and has achieved nearly two orders of magnitude higher efficiency than that of the conventional Mott detectors\cite{Okuda2009, Hillebrecht2002}.  On the other hand, the Mott polarimeters have been proven to be very stable and reliable in operation and therefore are the most widely used spin polarimeters, in spite of its inherent  low efficiency (generally ranging from $2 \times 10^{-5}$ to $1.6 \times 10^{-4}$).  A lot of progress has been made in improving the efficiency of the retarding Mott polarimeters by adopting good design of retarding optics with large collection solid angle of the scattering electrons.  The newly-developed so-called mini Mott polarimeters have efficiency better than $2\times 10^{-4}$ that are suitable for developing high-resolution SARPES systems\cite{SQiao1997,Hoesch2002,Neufeld2007}. Spin-resolved photoemission system equipped with Mott detectors is usually difficult to achieve high resolution. The energy resolution of spin-resolved ARPES by using Mott detectors was reported to be around 70 meV\cite{SQiao1997,Hoesch2002} and has been pushed to be around 8 meV very recently\cite{Souma2010}.  It is noteworthy that two-dimensional spin detection became available for hemispherical analyzers which drastically increase the spin detection efficiency\cite{Tusche2015}.

The utilization of laser light provides a good opportunity to enhance the performance of SARPES.  The idea is to make use of the extremely high photon flux of laser light to compensate the low efficiency  of the Mott spin detector while not losing the overall high energy resolution of the system. This is different from the synchrotron-based SARPES where the increase of photon flux has to be at the sacrifice of the light bandwidth, resulting in a poor energy resolution of 50$\sim$100 meV.  Based on this idea, a VUV laser-based SARPES with Mott spin detectors was developed in hope of achieving high energy resolution and efficiency.
%%Based on this idea, we started to develop VUV-laser based SARPES in 2008 with Mott spin detectors, in hope of achieving high energy resolution and efficiency, and got the SARPES system commissioned in 2012\cite{ZJXie2014a}.

Figure \ref{fig:SARPES} schematically shows the VUV laser-based SARPES which consists of two major parts:  the VUV laser optical system (see Fig. \ref{fig:Laser7eV}) and SARPES system. Two Mott detectors, which are placed perpendicular to each other, are put  onto the Scienta$-$Omicron  DA30 analyzer\cite{DA30} to fulfill the spin- and angle-resolved photoemission measurements. In order to perform ARPES and spin-resolved photoemission measurements at the same time, a portion of photoelectrons passes through two circular spin apertures with adjustable size. Each aperture is connected to a spin transfer lens, and the photoelectrons are transferred to the Mott detectors through the transfer lens for spin-resolved measurements. The photoelectrons are then accelerated  electrostatically to 25 keV through the input lens and hit the heavy element Gold (or Thorium) target. The intensity of the scattered electrons are counted by four channeltrons surrounding the target. If the incident photoelectrons are spin polarized, the intensities counted by a pair of channeltrons, for example, left and right channels in the inset of Fig. \ref{fig:SARPES}, will be different due to the strong spin-orbital interaction between the incident photoelectrons and the gold (thorium)  nucleus in the target. Such an intensity asymmetry from the left and right channels shown in the inset can be used to determine the spin component perpendicular to the scattering plane (paper plane for the inset). Two orthogonal spin components can be determined  by one Mott detector with four channels.   Two Mott detectors orthogonally placed as shown in Fig. \ref{fig:SARPES} can measure all three orthogonal spin components; one component is measured twice by both Mott detectors that can be used to calibrate the two Mott detectors.

Figure \ref{fig:SpinAu} shows the performance of the VUV laser-based SARPES system\cite{ZJXie2014a}. Benefitting from the intrinsic narrow lindwidth of the VUV laser light ($\sim$0.26 meV) as well as its super-high photon flux\cite{GDLiu2008}, we are able to achieve the best spin-resolved energy resolution of $\sim$2.5 meV (Fig. \ref{fig:SpinAu}c) that is comparable to the best result achieved so far for the spin resolved ARPES measurements (1.7 meV)\cite{KYaji2016}. Meanwhile the best angular resolution of our SARPES is around 0.3 degree and the corresponding momentum resolution can be further improved due to the lower photon energy of the VUV laser light (h$\nu$=6.994 eV).   There are  other outstanding capabilities of the VUV-laser based SARPES system: (1).  It is able to do spin- and angle-resolved photoemission measurements simultaneously.  This makes it possible to get all three variables of energy, momentum and spin at the same time. In addition, the momentum corresponding to the spin measurement can be determined precisely.   (2).  It can probe all three spin components of photoelectrons simultaneously.  (3). The laser light polarization can be tuned to have different linear or circular polarizations; (4). The utilization of the new DA30 analyzer makes it possible to cover two-dimensional momentum space without rotating the sample. This not only removes the mechanical uncertainty during the sample rotation, but also keeps the same polarization geometry during the SARPES measurements.

The spin-resolving capability of the SARPES was demonstrated by performing simultaneous  spin- and angle-resolved measurements on Au(111) surface states\cite{ZJXie2014a}.  Due to Rashba effects\cite{Rashba1960, Bychkov1984}, the surface state of Au(111) single crystals splits into two branches with well-defined spin polarization (Fig. \ref{fig:SpinAu}b)\cite{LaShell1996, Nicolay2001}. The spin polarized surface bands form two Fermi surface sheets with well-defined spin textures (Fig. \ref{fig:SpinAu}a): the inner cone has a left-handed chirality, whereas the outer one has a right-handed chirality\cite{Henk2004,Dil2009}.  In spin-resolved and angle-resolved mode, the surface band images (Fig.  \ref{fig:SpinAu}e) and spin-resolved energy distribution curves (EDCs) (Fig. \ref{fig:SpinAu}f) are obtained simultaneously by using the horizontal Mott detector. The band dispersion of the Au(111) surface state in Fig. \ref{fig:SpinAu}e is measured along the momentum direction depicted as a pink thick line in Fig. \ref{fig:SpinAu}a. The four spin-resolved EDCs in Fig. \ref{fig:SpinAu}f are recorded by the four channels of the horizontal Mott detector and correspond to the momentum point highlighted by the dashed blue line in Fig. \ref{fig:SpinAu}e. Due to much improved energy and angular resolutions, two peaks coming from the two spin polarized surface bands are well resolved in the raw data of the spin-resolved EDCs (Fig. \ref{fig:SpinAu}f). There is little difference between the two EDCs from the up (U) and down (D) channels, indicating a negligible spin polarization along the Y direction that is an out-of-plane spin component. On the other hand, the two EDCs from the left (L) and right (R) channels show obvious asymmetry in their intensity at two peak positions. Moreover, the relative intensity from these two channels is opposite for the two peaks: for the 0.16 eV binding energy peak associated with the outer cone, the L channel intensity (blue line) is higher than that of the R channel (red line);  for the 0.05 eV peak related with the inner cone, it is reversed. This indicates that there exists spin polarization along the vertical Z axis. The spin directions of the inner and outer cones lie in the plane of the Au(111) surface are opposite. These observations are consistent with previous results on Au(111)\cite{Dil2009,Henk2004} and demonstrate the spin-resolving capability of the SARPES system.

\subsection{VUV Laser-Based Time-of-Flight ARPES}
In the photoemission process, when light is incident on the sample, electrons are emitted in an entire 2$\pi$ solid angle. In the first generation of ARPES technique, the electron energy analyzer can only collect electrons emitted along one small solid angle  at one time, while the majority of photoelectrons along other angles, although they are already there,  are not measured and wasted.  The advent of the second generation electron energy analyzer can collect angles along one line in the real space, transforming the ARPES technique from zero-dimension angle coverage to one-dimensional angle coverage.  This is a big jump in the ARPES technique because it increases the data acquisition efficiency significantly.  The simultaneous collection of multiple angles along one line under the same condition fundamentally changed the game of ARPES.  New analysis methods are proposed such as the momentum distribution curves (MDCs)\cite{Valla1999}, in addition to the usual energy distribution curves (EDCs).  This is also a significant step to transform ARPES from a traditional band mapping tool into a probe of many-body effects.  However, even in this case, the majority of photoelectrons emitted along other angles are not measured and wasted.  It is natural to ask whether it is possible to cover  the two-dimensional angle space simultaneously.  The latest generation of angle-resolved time-of-flight (ARToF) electron energy analyzer is developed just for realizing this purpose. Many efforts have been put along this direction in the world\cite{Berntsen2011,YHWang2012, CLWang2016,YZhang2017}.

%%King2011,

A typical ARToF electron energy analyzer (e.g., ARToF 10k from Scienta-Omicron\cite{DA30}) consists of several cylindrical electrostatic lens and a delay-line detector (DLD),  as schematically shown in Fig. \ref{fig:ARTOF}.  When a laser light pulse hits on the sample surface at a time t$_0$,   electrons emitted within the acceptance angle of the first objective lens are all collected and imaged onto a three-dimensional position sensitive detector placed at the end of the lens set.  Electrons with different emission angles will follow different trajectories inside the lens.  When a photoelectron hits the detector,  a time t$_1$ will be recorded.  The corresponding kinetic energy of the electron is determined by the time it spends (t$_1$ - t$_0$)  along the particular trajectory. The delay-line detector can measure the position of the electron that hits the detector. The emission angle of the electron can be determined from the location on the DLD because there is a one-to-one correspondence map.  In this way, four dimensional data can be collected simultaneously, including the kinetic energy of electrons, the emission direction of the photoelectrons that can be defined by two angles, and the intensity of the photoelectrons.

Two kinds of VUV laser light sources are used for the ARToF-ARPES system.  The VUV laser system with a photon energy 6.994 eV follows the similar design of the VUV 6.994 eV laser system used in the laser-based ARPES system that is obtained through SHG of 355 nm photon from the nonlinear optical crystal KBBF\cite{GDLiu2008}.   The 355 nm (3.497 eV) 10 ps pumping laser light pulse comes from a Nd:YVO$_4$ laser light source.  It is focused onto the KBBF-PCT device inside the frequency doubling chamber.  After frequency doubling process, the diffused 177.3 nm (6.994 eV) VUV laser light is  focused onto the sample position in the ARPES measurement chamber. The beam size is tunable by choosing different focus lens.   Different photon polarizations (linear and circular) can be achieved with a half wave plate and a quarter wave plate.  The main difference between this 6.994 eV laser system with the previous one used in laser ARPES system\cite{GDLiu2008} is that the laser light source has a  low repetition rate between 0.2 MHz and 1 MHz.   Another  VUV laser light source with a photon energy 10.897 eV is equipped which is  based on the non-linear optical sum-frequency conversion process\cite{Lumeras}.

There are a number of advantages in the VUV laser-based ARToF-ARPES system. First, it measures two-dimensional momentum space simultaneously. Compared with the regular hemisphere analyzer that measures one line of momentum, the efficiency of momentum coverage is increased by more than 2 orders of magnitude.  Like the jump from the first generation one-angle-at-one-time to the second generation of one-line-at-a-time,  this jump to the third generation of one-area-at-a-time will bring about a fundamental change of the ARPES technique.  It makes it possible to observe the Fermi surface in real time that is impossible before. Since the two-dimensional angles are measured under the same condition, it will help to reveal fine electronic structure that are hard to resolve before.  Second, the VUV laser-based ARToF-ARPES has high energy and momentum resolutions. Third, the time-of-flight analyzer counts the number of electrons hitting the detector; this is different from the usual hemisphere analyzer where the photoelectron signal is first converted to photon signal by using a phosphor and then the intensity of the photons is recorded by a CCD camera.  The ARToF-ARPES measurements can give a low background, and suppress the signal nonlinearity effect that is usually encountered in a hemisphere analyzer.

To demonstrate the performance of the VUV laser ARToF-ARPES system,  Fig. \ref{fig:Eresolution}a shows the Fermi edge of a polycrystalline Au sample,  measured with the 6.994 eV laser light at a low temperature of $\sim$9.5 K which gives a  Fermi edge width of 3.37 meV.  After subtracting the temperature broadening,  the overall instrumental energy resolution of the 6.994 eV laser ARToF-ARPES system is better than 1 meV.  In Fig. \ref{fig:Eresolution}(b-e), the space charge effect was tested using the same polycrystalline Au measured with different powers and repetition rates of the laser light. An obvious space charge effect was observed when 0.1 mm spot size and 200 kHz were used (Fig. \ref{fig:Eresolution}b).   When the repetition rate was increased up to $\sim$1MHz, the space charge effect became nearly absent  (Fig. \ref{fig:Eresolution}c). Fig.  \ref{fig:FigNonLE} shows the 6.994 eV laser-based ARToF-ARPES measurements on an optimally-doped Bi$_{2}$Sr$_{2}$CaCu$_{2}$O$_{8+\delta}$ (Bi2212) superconductor measured at 15 K and Fig. \ref{fig:FigNonLE}a is the measured Fermi surface.  Due to the data continuity in the covered momentum space, one can extract high-quality band structure along any momentum cut (Fig. \ref{fig:FigNonLE}(f-j)). Such a capability of the data analysis is hard to be realized in the regular ARPES based on the hemispherical energy analyzer.   Non-linearity effect is examined   by taking the ARPES measurements at a low count rate (Fig. \ref{fig:FigNonLE}b) and a  high count rate (Fig. \ref{fig:FigNonLE}c) along the same momentum cut following a proposed  procedure\cite{Reber2014}.  The result indicates that there is nearly no nonlinearity effect in the ARToF-ARPES system (Fig. \ref{fig:FigNonLE}d). This is further confirmed by the same lineshape of two photoemission spectra measured using different count rate(Fig. \ref{fig:FigNonLE}e).

\subsection{Laser-Based Time-Resolved ARPES}
Ultrafast dynamics in solids, such as the photo-induced phase transitions\cite{Nasu2001,Takubo2005,Polli2007,MKLiu2012,Morrison2014,JDZhang2016,Donges2016}, photo-induced hidden phases or photo-enhanced density wave orders\cite{Kim2012,Stojchevska2014,Singer2016}, and especially the photo-induced possible superconductivity\cite{Fausti2011,WHu2014,Mitrano2016}, have attracted broad attention in condensed matter physics. In these ultrafast studies, ultrashort infrared laser light pulses, far infrared laser light pulses or even THz radiations are generally used as the pump source to create non-equilibrium states in materials. The photon source is commonly used as the probe because it is relatively easy to achieve ultrashort pulses (femtoseconds) to study the dynamics of electronic structure.  The recently-developed femtosecond electron pulses from accelerators\cite{Weathersby2015} and ultrashort X-ray pulses from free-electron lasers\cite{Hartmann2016}  bring new opportunities in probing the dynamics of the motion of atoms in real space. Among all the ultrafast probing techniques, the time-resolved ARPES (tr-ARPES) has opened a new window into directly probing the electronic dynamics in superconductors\cite{Perfetti2007,Smallwood2012b,Kummer2012,Rettig2012,LXYang2014,WTZhang2014,Rameau2014,SLYang2015}, topological insulators\cite{YHWang2012,Hajlaoui2013,Crepaldi2013,Sobota2014}, density wave systems\cite{Schmitt2008,Petersen2011,Rohwer2011,HYLiu2013,Ishizaka2011}, and other strongly correlated systems\cite{Cavalieri2007,Papalazarou2012,Ulstrup2015,ZSTao2016,Sterzi2016}.

 %%with resolution in momentum space

In a tr-ARPES system, as schematically shown in Fig. \ref{Fig7}a,  an infrared pump laser light pulse drives the sample into a non-equilibrium state. Subsequently,  an ultraviolet laser light is used to probe the  electrons which are usually captured by an electron energy analyzer in an ARPES setup.  The time resolution is achieved by varying the delay time between the pump and probe pulses.  For t$<$0,  the probe pulse arrives before the pump pulse, and thus corresponding to an equilibrium measurement. For t$>$0,  the probe pulse arrives after the pump pulse, and hence corresponding to a non-equilibrium measurement.

In tr-ARPES experiments, infrared laser light pulses (Ti: sapphire laser) are usually used as the pump, far infrared laser light pulses are used in a few experiments, and so far there is no report of THz radiation used as the pump yet. The tunability of the pump photon energy in the time-resolved ARPES is limited by the low energy conversion efficiency in the non-linear process in far infrared range.

In tr-ARPES experiments, high repetition rate of the probe pulse is necessary in order to photoemit reasonable number of electrons and, at the same time,  to reduce the space charge effect for achieving a reasonably high energy resolution. According to the generation mechanism, the probe photon sources in tr-ARPES can be divided into three classes. The first is from the frequency conversion in non-linear crystals, and the wavelength is in the ultraviolet
range with a photon energy usually no higher than 7 eV.  The tr-ARPES setup based on such a solid laser system is adapted by many groups because of the advantages of low cost, high stability, high energy resolution and easy operation, although there is a limitation on the accessible momentum space of a material\cite{Carpene2009,Smallwood2012a,YHWang2012,Faure2012,Ishida2014,SLYang2015}. A typical schematic of such a setup is shown in Fig. \ref{Fig7}b.  Laser light pulses are generated using a mode-locked Ti:sapphire oscillator with center wavelength around 836 nm. Two separated BBO nonlinear crystals are used to double the frequency of the output beam from the oscillator twice, achieving ultrashort ultraviolet probe pulses with a centre wavelength around 209 nm. The other branch of the beam (839 nm) directly outputted from the laser oscillator is used as the pump. The two beams are focused precisely at the same point on the sample in the ARPES chamber. Time resolution is achieved by precisely varying the optical path length of the pump beam using a fine linear translation stage. The overall time resolution of such setups can be better than 100 fs and the corresponding energy resolution is around 20 meV.

The second class of the probe source in the tr-ARPES is based on HHG process in gas phases, which are usually used to produce ultrashort laser light pulse with much shorter wavelength and shorter pulse length\cite{Siffalovic2001,Nugent2002,Mathias2007,Dakovski2010,Ishizaka2011,Wernet2011,Frietsch2013,HWang2015,Conta2016}. In addition to the broad bandwidth in frequency of the probe pulses due to the Fourier transfer limit, the low repetition rate of such a system also induces strong space charge effect, making it difficult to measure samples with a high energy resolution. Recently, a high repetition rate (250 kHz) HHG system was developed in tr-ARPES experiment,  but it has a very low probe photon flux because of the low three-order harmonic generation efficiency\cite{Cilento2016}. The third class of the probe source in the tr-ARPES is new and is under development in conjunction with  free-electron laser sources\cite{Pietzsch2008,Hellmann2012,Oloff2014}. Such laser source will have advantages of variable probe photon energy and pulse duration. However,
at present,  this technique is limited by the low repetition rate of the pulses and poor energy resolution.
%%this technique is also limited by the low repetition rate of the pulses because the energy resolution is important in most time-resolved ARPES studies in strongly correlated materials.

\section{Scientific Applications of Laser-Based ARPES Systems}

\subsection{New Coupling Mode and Extraction of the Eliashberg Functions}

The dramatic improvement of the ARPES technique has transformed it from a usual band mapping tool into a probe on the many-body effects in materials\cite{XJZhou2007}.  This is particularly important for understanding the anomalous normal state properties and the superconductivity mechanism of the high-temperature cuprate superconductors. The first indication of many-body effects in cuprates was observed in ARPES measurements on Bi2212 where a dispersion kink is identified along the nodal direction\cite{PVBogdanov,AKamiski,PJohnson,Lanzara2001,XJZhouNature}.  This was made possible mainly due to the advent of the second generation electron energy analyzer  that can measure multiple angles along a line simultaneously.  Since the many-body effects are subtle and hard to detect,  high resolution and high data statistics are required in the ARPES measurements. This is particularly important when quantitative analysis of the many-body effects are performed\cite{JRShi2004, XJZhou2005}.

The development of VUV laser-based ARPES with superior energy and momentum resolutions, as well as high data statistics due to high photon flux, provides a powerful tool to investigate the many-body effects in materials.  Fig.  \ref{fig:NewModeinBi2212} shows one example to demonstrate the power of the laser-based ARPES on this issue which  studied the nodal band renormalization in a Bi2212 superconductor\cite{WTZhang2008a}.  The original high precision ARPES data is shown in Fig.  \ref{fig:NewModeinBi2212}a and quantitative dispersion relation and the scattering rate are shown in  Fig.  \ref{fig:NewModeinBi2212}b.  Fig.  \ref{fig:NewModeinBi2212}c shows the effective real part of the electron self-energy at different measurement temperatures.  In addition to the well-known 70 meV mode-like feature that shows up very clearly in the measurements, there are new features emerging at high energies:  a dip-like feature at $\sim$115 meV and a hump structure at $\sim$150 meV, which are not resolved before.  They are only present in the superconducting state and their amplitude gets stronger with decreasing temperature.  Since the energy scale of phonons in Bi2212 is below 100 meV\cite{RJMcqueeney2001}, these new energy scales are unlikely   due to  electron coupling with phonons; some new mode couplings may exist in the superconducting state of Bi2212\cite{PCDai1999,HFFong2000,JHwang2007}.

One of the big challenges in the study of high-temperature superconductors is to understand the superconductivity mechanism. One critical issue concerns how the electrons interact and form Cooper pairs.  In the conventional superconductors, it was proposed by the BCS theory of superconductivity that the electron pairing is mediated by exchanging phonons.  The experimental validation of the BCS theory in the conventional superconductors is realized by the tunneling experiment\cite{Giaever}  and the subsequent elegant analyses\cite{WLMcMillan}.  The pairing Eliashberg function extracted from the tunneling data of Pb by direct inversion of the gap function exhibits a striking similarity to the phonon density of states measured directly from neutron scattering.  This provided a compelling evidence that the electrons in Pb form Cooper pairs by exchanging phonons.  It is natural to ask whether similar approach can be used in high temperature cuprate superconductors.  This turns out to be unfeasible because high T$_c$ superconductors show a {\it d}-wave superconducting order parameter that is different from the s-wave case in conventional superconductors like Pb. In this case, experimental tools that have momentum resolution have to be employed and ARPES turned out to be a promising technique to address the issues\cite{IVekhter}. Although the idea has been around since 2003,  no attempts have been successful because extremely high precision ARPES measurements are necessary for such analyses, and for a long time no ARPES data can satisfy such a demanding requirement.

The development of the VUV laser-ARPES provides an opportunity to address the important issue of pairing mechanism in high temperature cuprate superconductors by extracting the pairing Eliashberg functions in the superconducting state\cite{Bok2010}.  Since the pairing self-energy, and then the pairing Eliashberg function, are obtained from the signal difference between the superconducting state and the normal state, measurements of such a weak superconductivity-induced effect require high data precision (better than 1$\%$), in addition to high energy and momentum resolutions.   Such a stringent demand is recently met in the VUV laser-based ARPES,  the related results are summarized in Fig. \ref{fig:Eliashberg}.   From high resolution ARPES data measured below T$_c$ (Fig. \ref{fig:Eliashberg}a) and above T$_c$ (Fig. \ref{fig:Eliashberg}b),  with the analysis of  superconducting Green's functions,  the normal self-energy $\Sigma$ (Fig. \ref{fig:Eliashberg}c) and pairing self-enery $\phi$  (Fig. \ref{fig:Eliashberg}d and e) are successfully extracted.   By direct inversion of the Eliashberg equations, the normal Eliashberg function (${\cal E}_N$) and pairing Eliashberg function (${\cal P}_N$) are extracted (Fig. \ref{fig:Eliashberg}).  The extracted pairing Eliashberg function shows a peak at $\sim$50 meV, a flat featureless region extending to high energy, and a cut-off at high energy depending on the location of the momentum cuts. This is the first time that the normal Eliashberg function and the pairing Eliashberg function are extracted from ARPES measurements. They provide critical information on examining various theoretical models for understanding high-temperature superconductivity\cite{Bok2016}.

\subsection{Direct Observation of Spin-Orbital Locking in Topological Insulators}
Topological insulators have attracted much attention because of their unique electronic structure and spin texture, as well as their novel physical properties\cite{Hasan2010, SCZhang2011}. In its metallic surface state with helical spin texture,  the electron spin is locked to its crystal momentum\cite{Hasan2010, SCZhang2011}. Theoretical calculations also predict that the spin texture might be coupled with the orbital texture in topological insulators\cite{HJZhang2013}.   Such a new spin-orbital texture, if proved experimentally, can provide new insight on the spin texture of topological insulators and paves a way for light manipulation on spin textures in topological insulators.  To directly prove the possible spin-orbital locking in topological insulators, it is suggested that high resolution spin-resolved ARPES must be used.  In addition, the light must have tunable polarizations\cite{HJZhang2013}.

The VUV laser-based SARPES system can meet all the requirements for examing the spin-orbital locking in topological insulators\cite{ZJXie2014a}.  In order to do orbital-selective SARPES measurements, different linear polarization states of the incident VUV laser light with {\it s}- or {\it p}-polarization geometries were set and  different orbital textures of the Bi$_{2}$Se$_{3}$ Dirac surface state were observed by taking advantage of the photoemission matrix element effects.  The orbital textures of the Dirac cone detected under {\it s}- and {\it p}-polarization geometries are schematically shown in Fig. \ref{fig:SpinOrbital}e and f, respectively. The spin texture of the topological surface state of Bi$_{2}$Se$_{3}$ under different polarization geometries were further measured, as shown in Fig. \ref{fig:SpinOrbital}, which contains both the band mapping of the topological surface states (Fig. \ref{fig:SpinOrbital}c and d) and the spin-resolved photoemission spectra (Fig.  \ref{fig:SpinOrbital}a and b).  The spin-resolved photoemission spectra (energy distribution curves, EDCs) in Fig.  \ref{fig:SpinOrbital}a are measured under the {\it s} polarization geometry at five representative momentum points along the $\bar{\Gamma}\bar{K}$ cut as marked by the dashed lines in the band image measured from regular ARPES (Fig. \ref{fig:SpinOrbital}c). Likewise, spin-resolved EDCs in Fig. \ref{fig:SpinOrbital}b with the {\it p} polarization geometry are measured on five momentum points along the $\bar{\Gamma}\bar{K}$ cut as marked by the dashed lines in the band image in Fig. \ref{fig:SpinOrbital}d. At a given momentum point, the two EDCs represent the in-plane (sample surface plane) spin component along the vertical Z direction with up (red line plus circles) and down (blue line plus triangles) spin orientation (for definition of the spin direction, see Fig. \ref{fig:SpinAu}d ). By systematic observation of the momentum-dependent spin polarization of the Dirac cone in {\it s}- and {\it p}-polarization geometries, the spin texture is determined and the relationship between the spin and orbital textures is revealed, as summarized schematically in Fig. \ref{fig:SpinOrbital}e and f.  The spin textures observed in {\it s}- and {\it p}-polarization geometries are different. In particular, for {\it s}-polarization geometry, the upper Dirac cone and lower Dirac cone share the same chirality, which is distinct from the usual observation of spin texture in the topological surface state. On the other hand, in {\it p}-polarization geometry, the spin textures of the upper Dirac cone and lower one show opposite chirality, which is consistent with the total spin texture of the topological surface state. By switching the incident laser light from {\it p}- to {\it s}-polarization geometry, the spin texture for the upper Dirac cone changes from the left-handed to the right-handed chirality while it keeps the same for the lower Dirac cone.  These observations cannot be explained by the usual spin picture in topological insulators, but are fully consistent with the predictions in terms of the spin-orbital locking.  These results thus provide a strong experimental evidence for the orbital-selective spin texture in the Bi$_{2}$Se$_{3}$ topological insulator\cite{ZJXie2014a}.

\subsection{Novel Electronic Structure of WTe$_{2}$ and ZrTe$_{5}$ Revealed by VUV Laser-Based ARToF-ARPES}
\subsubsection{Complete electronic structure and topological nature of WTe$_2$}

WTe$_2$ has been well-known for its manifestation of extremely large magnetoresistance that is proposed to be due to the compensation of electrons and holes in the material\cite{1AMN_N_2004}.  A complete understanding of its electronic structure is a prerequisite to pin down the origin of the anomalous transport properties. However, due to the existence of multiple pockets in a limited momentum space, the full electronic structure of WTe$_2$ remains controversial.  Lately, WTe$_2$ has ignited another surge of excitement because it is theoretically predicted to be the first material candidate that may realize type-II Weyl state\cite{A.Soluyanov}.  However, experimental evidence on the identification of type II Weyl Fermions in WTe$_2$ is still lacking.

Taking advantage of the VUV laser-based ARToF-ARPES system with superior instrumental resolution, a complete picture of the electronic structure of WTe$_{2}$ has been resolved\cite{CLWang2016}.  Fig. \ref{FigWTe2}  shows the measured Fermi surface of WTe$_2$ where the existence of a surface state is clearly identified.  High temperature ARPES measurements make it possible to reveal electronic states above the Fermi level where the Weyl points are predicted to be located. The observed connection of the surface state with the bulk bands, its momentum evolution, its momentum and energy locations, are all in good agreement with the calculated band structure.  These results provide electronic signatures that are consistent with the type II Weyl state in WTe$_2$ although further efforts are needed to fully prove the realization of type II Weyl state in WTe$_2$.  The first-time complete electronic structure of WTe$_2$, including all the Fermi pockets and near-E$_F$ energy bands also makes it possible to determine accurately the electron and hole concentrations and their temperature dependence\cite{CLWang2017}.  With increasing temperature, the overall electron concentration increases while the total hole concentration decreases. It indicates that, the electron-hole compensation, if exists, can only occur in a narrow temperature range and in most of the temperature range there is an electron-hole imbalance. These results are not consistent with the perfect electron-hole compensation picture that is commonly considered to be the cause of the unusual magnetoresistance in WTe$_2$. They provide new insight on understanding the origin of the unusual magnetoresistance in WTe$_2$\cite{CLWang2016,CLWang2017}.

%%\ref{fig:FigWTe2}

\subsubsection{Temperature-Induced Lifshitz Transition and Topological Nature in ZrTe$_5$}

The topological materials have attracted much attention recently for their unique electronic structure, spin texture and peculiar physical properties. While three-dimensional topological insulators are becoming abundant, two-dimensional topological insulators remain rare, particularly in natural materials.  ZrTe$_5$ has host a long-standing puzzle on its anomalous transport properties manifested by its unusual resistivity peak and a sign reversal of the Hall coefficient and thermopower;  the underlying origin remaining elusive.   Lately, ZrTe$_5$ has ignited renewed interest because it is predicted that single-layer ZrTe$_5$ is a two-dimensional topological insulator and there is possibly a topological phase transition in bulk ZrTe$_5$\cite{HW_PRX_2014}. However, the topological nature of ZrTe$_5$ remains under a heated debate.

By using the VUV laser-based ARToF-ARPES system,  high-resolution ARPES measurements are performed on the electronic structure and its detailed temperature evolution of ZrTe$_5$\cite{YZhang2017}.  As displayed in Fig. \ref{FigZrTe5}, the electronic property of ZrTe$_5$ is dominated by two branches of nearly-linear dispersion bands with a gap between them at the Brillouin zone center. The overall electronic structure exhibits a dramatic temperature dependence; it evolves from a p-type semimetal with a hole-like Fermi pocket at high temperature, to a semiconductor around $\sim$135 K where its resistivity exhibits a peak, to an n-type semimetal with an electron-like Fermi pocket at low temperature. These results provide direct electronic evidence on the temperature-induced Lifshitz transition in ZrTe$_5$, and a natural understanding on the underlying origin of the transport anomaly at $\sim$135 K. In addition, one-dimensional-like electronic features are observed from the edges of the cracked ZrTe$_5$ samples implying that ZrTe$_5$ is a weak topological insulator\cite{YZhang2017}.

\subsection{Time-Resolved ARPES on High-Temperature Cuprate Superconductors and Topological Insulators}
Time-resolved ARPES has been widely used in studying the momentum-resolved electronic dynamics in high-temperature superconductors, topological insulators, density wave systems, and other strongly-correlated systems. After photon excitation, the electron-boson coupling, specific mode vibration, the superconducting gap, and so on can be tracked by tr-ARPES. Here we highlight two applications of tr-ARPES in the study of ultrafast electronic dynamics in the high-temperature superconductors and topological insulators.
	
It remains unclear how electrons bind into Cooper pairs in high-temperature cuprate superconductor and what the origin of the abnormal pseudogap phase is in the underdoped region. Tr-ARPES provides a unique approach to probe the non-equilibrium quasiparticle dynamics\cite{Perfetti2007,Cortes2011,Smallwood2012b,SLYang2015}, non-equilibrium phase diagram\cite{WTZhang2013}, many-body problem\cite{WTZhang2014,Rameau2014}, energy gap dynamics\cite{Smallwood2014}, and so on. The many-body interactions is important in understanding the superconductivity mechanism. Tr-ARPES has the capabilities of tracking the electron-boson interaction dynamics and the non-equilibrium superconducting gap simultaneously, providing information to clarify the role of electron-boson coupling in generating superconductivity.

Figure \ref{Fig13} shows an example of  such a study along this direction\cite{WTZhang2014}. In the optimally-doped Bi2212, the nodal kink around 70 meV  signifies electrons coupled with some bosons at this energy scale. Far below the superconducting transition temperature T$_c$, intense ultrashort laser light pulse weakens the kink strength apparently (Figs. \ref{Fig13}a and b).  This can be further confirmed by the extracted effective real part of the electron self-energy shown in the lower panel of Fig.  \ref{Fig13}c. By simultaneously tracking the superconducting energy gap, it is found that the photo-induced change of the electron-boson coupling strength is proportional to the change of the superconducting energy gap, while above T$_c$ when the superconducting gap fully closes, pumping photons has a negligible effect on the kink (upper panel in Fig.  \ref{Fig13}c). Similar negligible photo-induced change of the kink structure is also observed in the normal state of a heavily overdoped Bi2201 sample at very low temperature\cite{WTZhang2014}. These findings establish a strong correlation between the nodal electron-boson coupling and the superconductivity in high temperature cuprate superconductors, and demonstrate that tr-ARPES is an emerging powerful tool to address non-equilibrium self-energy dynamics and many-body interactions in materials.

In topological insulators, such as Bi$_2$Se$_3$,  there is no direct electron-hole recombination because of the opposite spin texture at the same momentum in the Dirac dispersion.  It is therefore necessary to use momentum-resolved technique, such as  tr-ARPES, to study the manipulation of the exotic surface states by ultrashort laser light pulse\cite{Sobota2012,YHWang2012,YHWang2013,Sobota2013}.

When moving in a periodic potential (crystal lattice), electrons' energy locks on to their momentum periodically in a parabolic form and are gapped at the boundary of the Brillouin zone;  this is named Bloch states.  Under an intense electromagnetic field (usually can be created by an ultrashort intense laser light pulse), electrons move in solids periodically in both space and time, and this is called Floquet-Bloch states. When pumping Bi$_2$Se$_3$ with linear polarized laser light pulses, the periodic duplication of dispersion in energy is demonstrated by tr-ARPES as shown in Fig. \ref{Fig14}a. The energy gap can be resolved at the band crosses away from the Dirac point (Figs. \ref{Fig14}b and c). Further studies show that using circular polarized laser light pulse, a gapped Dirac cone can be observed, indicating the broken time-reversal symmetry induced by ultrashort laser light pulse. These observations provide strong evidence of the Floquet-Bloch bands in solids and pave the way for optical manipulation of quantum states in matter\cite{YHWang2013}.

Except for the novel study in optical control of the topological surface states, the non-equilibrium quasiparticle dynamics induced by ultrashort laser light pulses in topological insulators can also be tracked by tr-ARPES.  Fig. \ref{Fig14}d shows the non-equilibrium quasiparticle oscillations in both bulk and surface electronic states which are apparently different from each other. This indicates that there are different mechanisms in controlling the dynamics of photo-induced non-equilibrium quasiparticles in bulk and surface electronic states. Results of Fourier transforms confirm the observations and reveal an additional vibration mode at 2.05 THz only in the surface states, while in both the bulk and surface states there is another mode at a higher frequency (2.23 THz) (Fig. \ref{Fig14}(e)). The 2.23 THz vibration is attributed to an A$_{1g}$ phonon mode and the 2.05 THz vibration is due to the termination of the crystal and thus reduction of the interlayer van der Waals forces, which serve as restorative forces for out-of-plane lattice distortions\cite{Sobota2014}. The study showcases the power of tr-ARPES to resolve distinct modes coupling to individual bands.

\section{Outlook and Perspective}

ARPES has played a key role in studying the novel quantum materials in condensed matter physics.  There is no doubt that it will continue to play a leading role in the years to come. Many prominent physics problems ask for further improvement of the ARPES capabilities.  The emergence of laser-ARPES in the last ten years has created a new path for endowing ARPES with unique and superior performances.  It has already demonstrated its power in studying quantum materials like high temperature superconductors and topological materials.  There is still plenty of room for laser-ARPES to further enhance its capabilities.  We will expect its continuous improvement along the following directions:

1.  Laser light sources with higher photon energy, narrower bandwidth, and shorter pulse duration.

As we have seen, the laser photon energy for high-resolution ARPES has steadily increased from 6 eV, to 7 eV, to 11 eV in the last ten years.  To cover a large momentum space, it is important to further increase the laser photon energy, while keeping its unique characteristics like narrow bandwidth and extremely high photon flux.  There have been big efforts along this direction. High order harmonic generation is a promising approach to get higher laser photon energy, and further work is needed to increase its repetition rate and average power in order to be used for high resolution ARPES.

There is still room to further improve the energy resolution of the ARPES technique.  For example,  measurement of the superconducting gap in superconductors with lower T$_c$ such as Sr$_2$RuO$_4$ with a T$_c$ at $\sim$1 K and CeCoIn$_5$ with a T$_c$ at 2.3 K definitely requires an energy resolution much better than 1 meV.  Next generation ARPES should improve the  best energy resolution from 1 meV to the order of  $\mu$eV.  This requires laser light sources to have much narrower bandwidth.   Since photoemission process is closely related to the space charge effect that may deteriorate the energy resolution,  the ultimate energy resolution may be limited by the space charge effect, which becomes particularly serious in laser-ARPES because of  laser light's  high photon flux and relatively small beam size.  To overcome the space charge effect,  it is desirable to increase the repetition rate of the laser light sources in order to reduce the number of photons in one pulse.   Development of continuous-wave laser light will be most ideal because in principle it does not produce space charge effect\cite{Scholz2013,Tamai2013,Taniuchi2015}.  The continuous-wave laser light may become particularly useful for spin-resolved ARPES where high photon flux is necessary in order to compensate the low efficiency of spin detectors which will inevitably causes strong space charge effect if it is a pulsed laser light source.

Spatially-resolved ARPES will play an important role in studying inhomogeneous systems.   Focusing laser light to microns or sub-microns will be beneficial to perform spatially-resolved ARPES. For materials that are spatially inhomogeneous  or new materials that are hard to get big enough single crystals, in order to detect their intrinsic electronic structure, spatially-resolved ARPES is necessary. With the micron-scale or even nanoscale spot size, spatially-resolved ARPES will play an irreplaceable role in obtaining information about phase separation or multiple domains of materials.

Tunable laser light source that can cover a large range of photon energy needs to be developed.  This is helpful in dealing with the photoemission matrix element effects and increasing the variety of materials that can be measured. Furthermore,  tunable laser light is very important for  measuring materials with strong three-dimensionality of electronic structure.

Time-resolved ARPES is a powerful tool for investigating electron dynamics and non-equilibrium state.   For strongly correlated systems, such as high temperature superconductors, electronic dynamics happens with an energy scale on the order of meV and a time scale of femtosecond.  The development of laser light source that can produce very short pulse and high energy resolution will promote time-resolved ARPES applications.   High repetition rate, relatively narrow bandwidth, and tunability of the probe pulse are important in time-resolved ARPES studies. Also in future time-resolved APRES, developing strong tunable pump pulses from violet to far infrared, even to THz range, which can be used to coherently excite the low energy modes such as the lattice vibrations in materials, will be important in controlling the electronic structure in materials by strong laser light pulses.

With the fast development of free electron lasers, and considering their extremely high photo flux, ultrashort pulse width and spatial and temporal coherence, their extensive applications in photoemission is highly expected. Currently, the time-resolved photoemission experiments using free electron lasers face challenges of  very strong space charge effect, poor energy resolution and insufficient statistics due to very low repetition rate and too high pulse intensity. These issues can be addressed by significantly increasing the repetition rate up to $\sim$100 kHz - $\sim$MHz and manipulating the photon pulse energy and bandwidth. In this regard, LCLS-II and European XFEL employing superconducting RF accelerators may provide some unique opportunities in time-resolved photoemission techniques.

2. Improvement of the photoelectron detection technique

We have seen the dramatic improvement of the electron energy analyzer from the first generation to measure one angle at one time, to the second generation  to measure multiple angles along one line, to the present third generation to measure a two-dimensional area of angles. The detection of photoelectrons covering full 2$\pi$ solid angle at one time is ideal. The technique is under development and should be implemented in all the electron energy analyzers in the future. In the mean time, the energy resolution has also been improved from 10 meV to 1 meV to nowadays 0.1 meV.  Further increase of the energy resolution of the ARPES system to $\mu$eV level asks the analyzers to have even better energy resolution.  In the spin-resolved ARPES technique, concurrent detection of spin state during the simultaneous ARPES measurement of the two-dimensional momentum space represents a future direction to develop.

3.  Improvement of the sample environment

The utilization of laser light source in ARPES has resulted in super-high energy resolution better than  1 meV\cite{GDLiu2008,Kiss2008}.  However, the sample temperature will also cause spectral broadening.   For most of ARPES systems at work,  the lowest sample temperature is usually about 10 K which corresponds to  about 3.7 meV Fermi broadening.  This is significantly larger than 0.26 meV bandwidth of the VUV laser light source.  To take full advantage of the super-high energy resolution of laser-ARPES, it is imperative to lower the sample temperature to below 1 K or even lower.  Ultra-low sample temperature is also necessary for studying materials with low temperature scales, like superconductors with low T$_c$.  Efforts have been put into developing new cryostats with ultra-low temperature\cite{BorisenkoCubic}.   A VUV laser-based ARPES system with ultra-low sample temperature (below 1 K) has been developed in the institute of physics, Beijing\cite{http12}.

4. Combination of  ARPES with other techniques

For some deep physics problems, one needs to get information about various aspects in order to have a complete picture. Therefore,  integration of various photoemission techniques, such as angle-resolved photoemission, spin-resolved photoemission, spatially-resolved photoemission, and time-resolved photoemission,  in one system is a good option.  Also ARPES can be integrated with other techniques such as structural analysis tools to obtain both electronic structure and lattice and magnetic structures simultaneously.   Last but not least, because ARPES can only work on a clean and flat surface,  it becomes very fruitful to integrate sample preparation into the ARPES system. Combining ARPES with {\it in situ} sample preparation like molecular beam epitaxy (MBE) and  {\it in situ} sample characterization like scanning tunneling microscope (STM)  has produced significant results and will become popular in the future.

With further development of laser-ARPES and its related techniques,  we believe it will play more and more important role in studying novel quantum materials in the near future.

\bibliographystyle{unsrt}
%\bibliographystyle{abbrv}
%\bibliographystyle{alpha}

%\bibliography{LaserARPESRef}
%\bibliography{Test2017}

%%\vspace{3mm}
%%\noindent {\bf Supplementary Information}  is linked to the online version of the paper.

\vspace{3mm}

\noindent {\bf Acknowledgements} We acknowledge collaborations with Prof. Zuyan Xu's group and Prof. Chuangtian Chen's group in developing laser-based photoemission systems over many years,  and Prof. Chandra Varma and Prof. Han-Yong Choi for collaborations in data analysis and discussions. We acknowledge permission of using the FEL sketch provided by Siegfried Schreiber in DESY - FLASH. XJZ thanks the funding support by the National Key Research and Development Program of China (2016YFA0300300), the National Science Foundation of China (11334010),  the Strategic Priority Research Program (B) of the Chinese Academy of Sciences (Grant No. XDB07020300),and .  Wentao Zhang was sponsored by Shanghai Pujiang Program.
\vspace{3mm}

%%\noindent {\bf Author Contributions}
 %% wrote the paper.

%%\vspace{3mm}
%%\noindent {\bf Author Information} Correspondence and requests for materials should be addressed to X.J.Z. (XJZhou@aphy.iphy.ac.cn).

\newpage
%%Begin Figures
%%Introduction

%\begin{figure}
%\centering
%\includegraphics[width=1.0\textwidth]{IOPLaserARPES.png}
%\caption{\label{ARPES}  Layout of four VUV laser-based ARPES systems developed in the Institute of Physics, Chinese Academy of Sciences, Beijing. }
%\label{fig:IOPLaserARPES}
%\end{figure}

%%\begin{figure}
%%\centering
%%\includegraphics[width=1.0\textwidth]{FigPESDOS.png}
%%\caption{\label{PESDOS} Relation between energy levels inside a solid and the photoelectron kinetic energy distribution curve produced by photons of energy $\hbar\nu$.}
%%\label{fig:PESDOS}
%%\end{figure}

%%LASER
%\begin{figure}
%\begin{center}
%\includegraphics[width=1.0\columnwidth,angle=0]{FigLaserlistCrop}
%\end{center}
%\caption{List of some typical lasers used in ARPES.  (a).  First high resolution VUV-laser ARPES; (b).  First high resolution UV-laser ARPES; (c).  The highest resolution VUV-laser ARPES; (d).  Tunable laser ARPES; (e).  First CW UV laser ARPES; (f).  First CW deep-UV laser for ARPES; (g).  Quasi-CW laser tr-ARPES; (h).  Good energy and time resolution compromised tr-ARPES; (i).  Very high time-resolution tr-ARPES; (j).  First high resolution HHG-laser ARPES; (k). First high resolution resonant type HHG-laser ARPES}
%\label{fig:Laserlist}
%\end{figure}
\begin{table*}[!hbp]
\tiny
\centering
\caption{Some typical laser light sources used in ARPES}
\begin{tabular}{|c|c|c|c|c|c|c|c|c|c|c|}
\toprule[1pt]
\hline
 \makecell[cc]{Laser\\Catagory}   & Generation & Application & \makecell[cc]{Pho. Energy\\(eV)} & \makecell[cc]{Pulse width\\(ps,fs)} & \makecell[cc]{Rep. Rate\\(kHz,MHz)} & \makecell[cc]{Max.Pho.flux\\(photons/s)} & \makecell[cc]{Energy Res.\\(meV)} & \makecell[cc]{Tem. Res.\\(fs)} & Refs & Remarks\\\midrule[1pt]
\hline
\multirow{4}{*}{Quasi-CW} & \multirow{4}{*}{\makecell[cc]{NLO crystal\\SFG+SHG}} & \multirow{4}{*}{\makecell[cc]{High-Res\\ARPES}}
%\cline{4-11}
& 7 & $\sim$10ps & 80MHz & 1.5$\times10^{15}$ & 0.26meV & / & G.D.Liu et al.\cite{GDLiu2008} & a \\
\cline{4-11}
 \multirow{4}{*}{} & \multirow{4}{*}{} & \multirow{4}{*}{} & 6 & $\sim$70fs(seed) & 100MHz & $\sim10^{15}$ & 4.7meV & / & J.D.Koralek et al.\cite{Koralek2007} & b \\
\cline{4-11}
\multirow{4}{*}{} & \multirow{4}{*}{} & \multirow{4}{*}{} & 7 & $\sim$10 ps & 120MHz & unknown & 0.025meV & / & K.Okazaki et al.\cite{Okazaki2012} & c \\
\cline{4-11}
\multirow{4}{*}{} & \multirow{4}{*}{} & \multirow{4}{*}{} & 5.3-7 & 5 ps & 76MHz & $\sim10^{14}$ & unsecified & / & R.Jiang et al.\cite{RJiang2014} & d \\
\hline
\multirow{2}{*}{CW} & \multirow{2}{*}{\makecell[cc]{NLO crystal\\SFG+SHG}} & \multirow{2}{*}{\makecell[cc]{High-Res\\ARPES}} & 6.05 & infinite & infinite & 1$\times10^{15}$ & 0.01 & / & A.Tamai et al.\cite{Tamai2013} & e \\
\cline{4-11}
\multirow{2}{*}{} & \multirow{2}{*}{} & \multirow{2}{*}{} & 6.49 & infinite & infinite & 1.25$\times10^{15}$ & $\sim 10^{-7}$meV & / & M. Scholz et al.\cite{Scholz2012} & f \\
\hline
\multirow{3}{*}{\makecell[cc]{Pulsed\\Laser}} & \multirow{3}{*}{\makecell[cc]{NLO crystal\\SFG+SHG}} & \multirow{3}{*}{Tr-ARPES} & 1.5,6 & 50 fs,160 fs & 80MHz & unspecified & $<$22meV & 163 fs & J.A.Sobota et al.\cite{Sobota2012} & g \\
\cline{4-11}
\multirow{3}{*}{} & \multirow{3}{*}{} & \multirow{3}{*}{} & 1.5,6.04 & 35 fs,55 fs & 250kHz & $\sim10^{13}$ & 40meV & 65 fs & J.Faure et al.\cite{Faure2012} &   \\
\cline{4-11}
\multirow{3}{*}{} & \multirow{3}{*}{} & \multirow{3}{*}{} & 1.48,5.92 & 170 fs,- & 250kHz & unspecified & $\geq$ 10.5meV & $\geq$240 fs & Y.Ishida et al.\cite{Ishida2014} &   \\
\hline
\multirow{3}{*}{HHG} & \multirow{3}{*}{\makecell[cc]{Noble gas\\HHG}} & \multirow{3}{*}{Tr-ARPES} & 1.58,15-40 & 40 fs,100 fs & 10kHz & 3.6$\times10^{17}$ & 90meV@35.6eV & 125 fs & B.Frietsch et al.\cite{Frietsch2013} &  h \\
\cline{4-11}
\multirow{3}{*}{} & \multirow{3}{*}{} & \multirow{3}{*}{} & 1.6,22.1 & 30fs,11 fs & 10kHz & unspecified & 170meV & 13 fs & G.Rohde et al.\cite{Rohde2016} &  i \\
\cline{4-11}
\multirow{3}{*}{} & \multirow{3}{*}{} & \multirow{3}{*}{} & 1.57,20.4 & 30 fs & 1kHz & unspecified & unspecified & 30 fs & J.C.Petersen et al.\cite{Petersen2011} &  \\
\hline
\multirow{2}{*}{\makecell[cc]{Mod./Reson.\\type HHG}} & \multirow{2}{*}{\makecell[cc]{Mixed rare\\gas}} & \multirow{2}{*}{\makecell[cc]{High-Res\\ARPES}} & 10.5 & 10 ps,& 0.2-8MHz & 9$\times10^{12}$ & $<1meV$ & / & M.H.Berntsen et al.\cite{Berntsen2011} &  j \\
\cline{4-11}
\multirow{2}{*}{} & \multirow{2}{*}{} & \multirow{2}{*}{} & 10.9 & 100 ps,& 1-20MHz & $\times10^{13}$ & $<2meV$ & / & Yu He et al.\cite{YHe2016} &  k \\
\hline
FEL & Long Undulator& Tr-ARPES & 26-300 & 30-150 ps & $<$10Hz& Very high & 300 meV& 700 fs & S Hellmann et al.\cite{Hellmann2012} &\\
\hline
\end{tabular}
%\normalsize

{(a).  First high resolution VUV-laser ARPES. (b).  First high resolution UV-laser ARPES. (c).  The highest resolution VUV-laser ARPES. (d).  Tunable laser ARPES. (e).  First CW UV laser ARPES. (f).  First CW deep-UV laser for ARPES. (g).  Quasi-CW laser tr-ARPES. (h).  Good energy and time resolution compromised tr-ARPES. (i).  Very high time-resolution tr-ARPES. (j).  First high resolution HHG-laser ARPES. (k). First high resolution resonant type HHG-laser ARPES.}
\label{fig:Laserlist}
\end{table*}
%\caption{List of some typical laser light sources used in ARPES.  (a).  First high resolution VUV-laser ARPES; (b).  First high resolution UV-laser ARPES; (c).  The highest resolution VUV-laser ARPES; (d).  Tunable laser ARPES; (e).  First CW UV laser ARPES; (f).  First CW deep-UV laser for ARPES; (g).  Quasi-CW laser tr-ARPES; (h).  Good energy and time resolution compromised tr-ARPES; (i).  Very high time-resolution tr-ARPES; (j).  First high resolution HHG-laser ARPES; (k). First high resolution resonant type HHG-laser ARPES}

%%\begin{figure}[tbp]
%%\begin{center}
%%\includegraphics[width=1.0\columnwidth,angle=0]{Fig3Step}
%%\end{center}
%%\caption{{\bf The 3 step model for the HHG process.} (a) Ionization of bonded electron; (b) Acceleration and reversal of electron; (c) Recombination of electron with the ion and emission of high energy photon.}
%%\label{fig:3Step}
%%\end{figure}

\begin{figure}
\begin{center}
\includegraphics[width=1.0\columnwidth,angle=0]{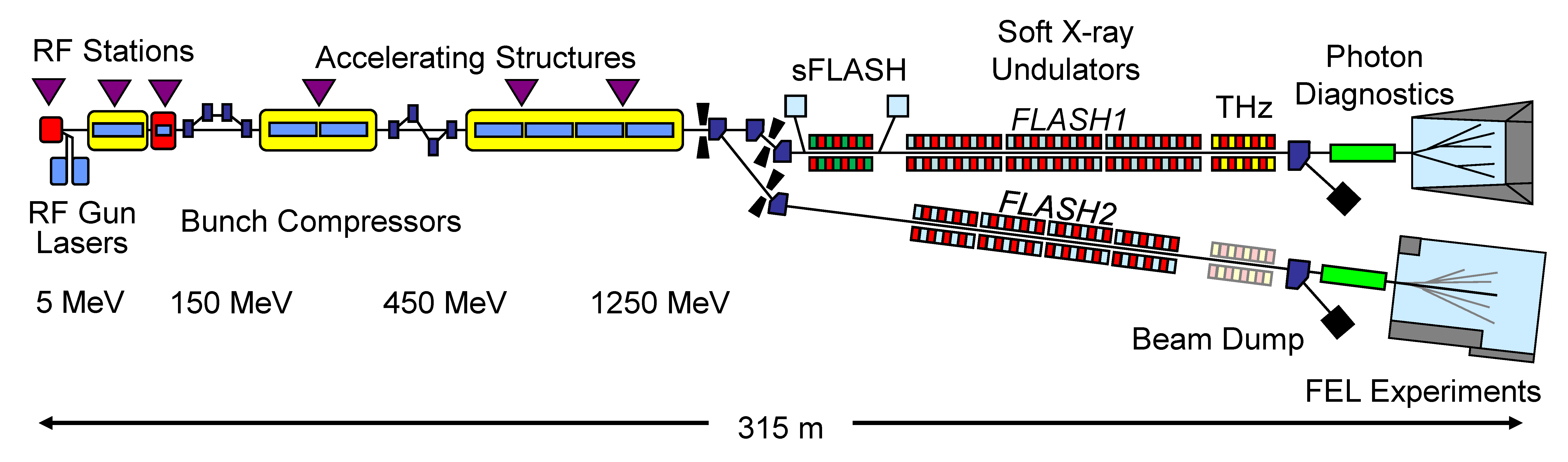}
\end{center}
\caption{A schematic layout of the first XUV and soft X-ray FEL apparatus FLASH\cite{http1}.}
\label{fig:FEL}
\end{figure}

\begin{figure}
\begin{center}
\includegraphics[width=1.0\columnwidth,angle=0]{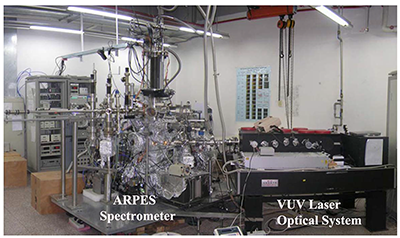}
\end{center}
\caption{A photograph of the VUV laser-based ARPES system.}
\label{fig:LaserARPES}
\end{figure}

\begin{figure}
\begin{center}
\includegraphics[width=1.0\columnwidth,angle=0]{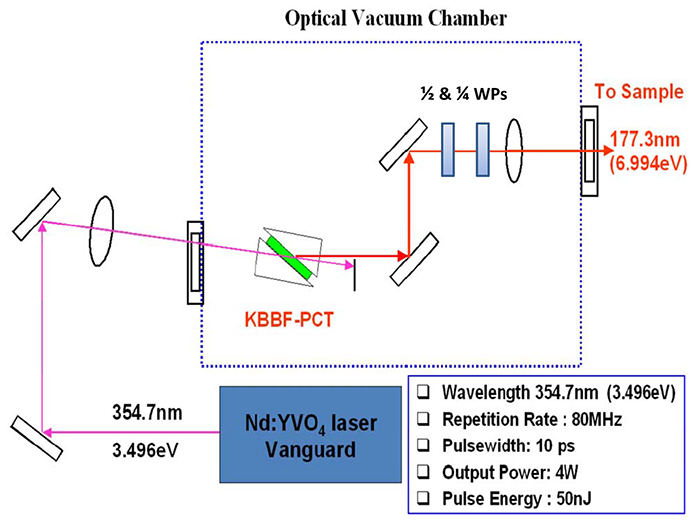}
\end{center}
\caption{A schematic layout of a VUV laser optical system\cite{GDLiu2008}. }
\label{fig:Laser7eV}
\end{figure}

\begin{figure}[tbp]
\begin{center}
\includegraphics[width=1.0\columnwidth,angle=0]{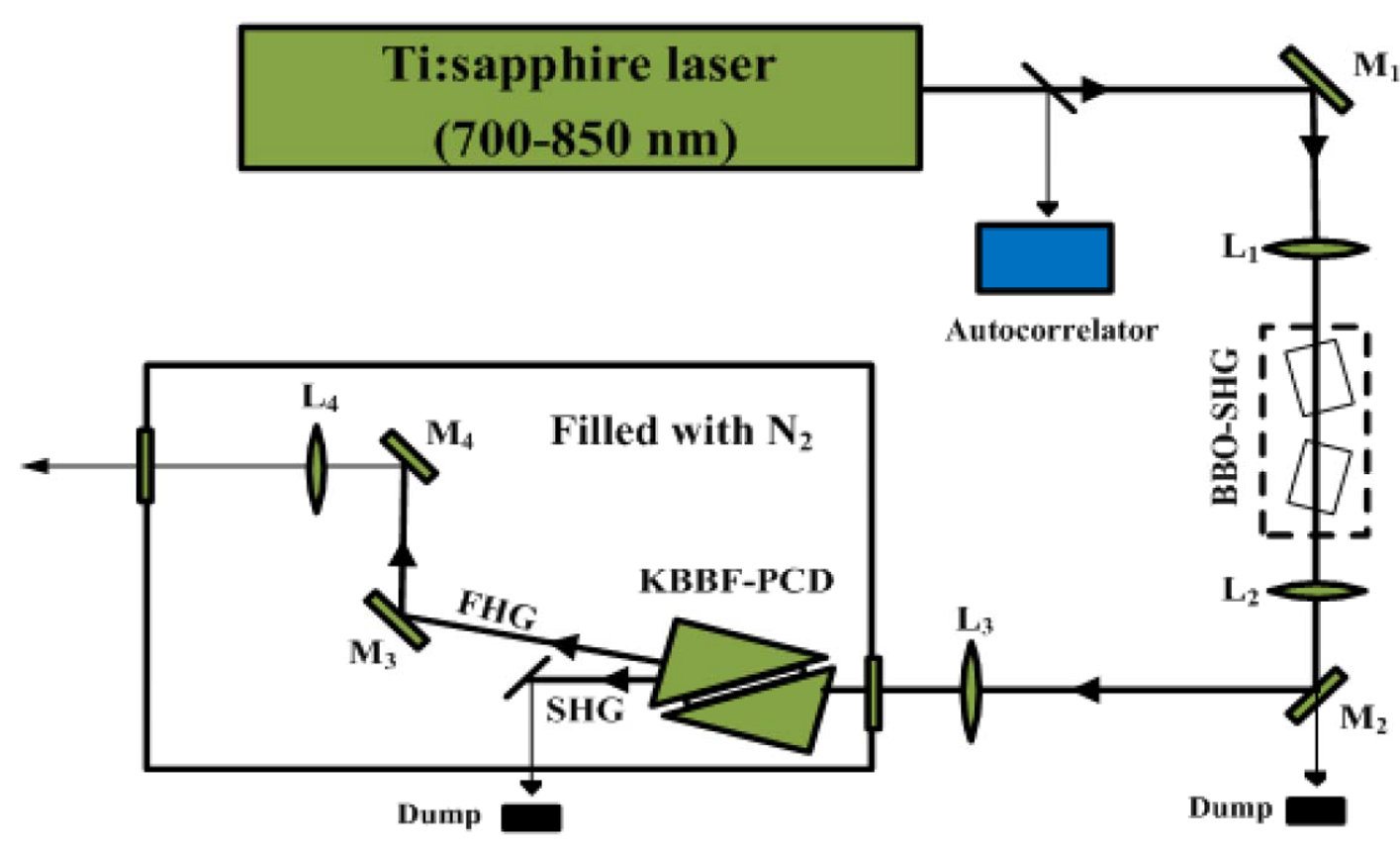}
\end{center}
\caption{A schematic layout of a tunable and deep-UV laser optical system\cite{SJZhang2014}.}
\label{fig:Tunable}
\end{figure}

%%Development of SARPES
\begin{figure}
\centering
\includegraphics[width=1.0\textwidth]{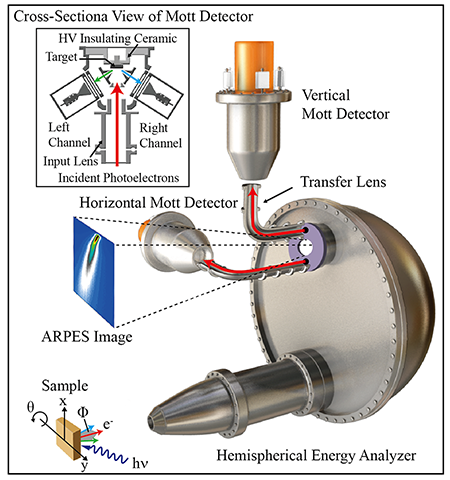}
\caption{ Working principle of a spin-resolved ARPES equipped with two  Mott type spin detectors. The inset is the cross-sectional view of the Mott detector. }
\label{fig:SARPES}
\end{figure}

\begin{figure}
\centering
\includegraphics[width=1.0\textwidth]{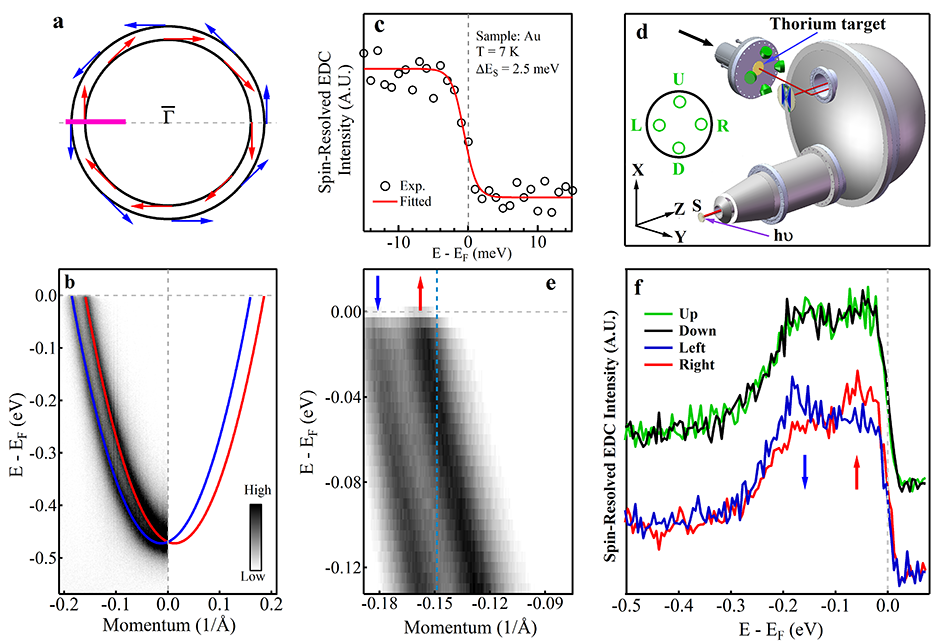}
\caption{\label{SpinAu} The spin-resolved ARPES measurements on the Au(111) surface state\cite{ZJXie2014a}.  {a}. Schematic Fermi surface and associated spin texture of the Au(111) surface state. {b}. Schematic band structure of the Au(111) surface state along the momentum cut shown in {a} as a dashed line. It is overlaid on the band structure measured by regular ARPES (left half of the momentum range).  {c}. Spin-resolved energy resolution test obtained by measuring the Fermi edge of a clean polycrystalline Au at 7 K. The measured data (open circles) are fitted by the Fermi distribution function (red solid line) and the overall fitted line width is 3.52 meV. By removing the thermal broadening, an instrumental spin-resolved energy resolution of 2.5 meV is obtained.  {d}. Schematic layout for a SARPES system which combines a Scienta R4000 electron energy analyzer with a Mott-type spin detector.  {e} and {f} show, respectively, the band structure image and the four spin-resolved EDCs of the Au(111) surface state obtained simultaneously by the SARPES system.  The corresponding momentum cut for the image ({e}) is shown in (a) as a pink thick line.  The corresponding momentum point for the spin-resolved EDCs (f) is shown in (e) as a dashed blue line.}
\label{fig:SpinAu}
\end{figure}

%%Development of ARTOF
\begin{figure}
\centering\includegraphics[width=1\columnwidth]{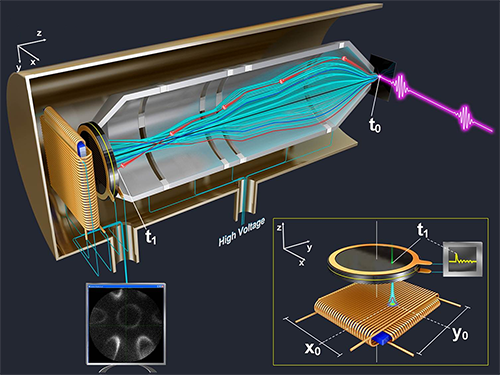}
\caption{Schematic 3D drawing of a 7 eV VUV laser-based ARToF-ARPES system. The analyzer consists of an electrostatic lens system and a MCP/DLD detector attached at the end of the lens. The light-blue lines represent the photoelectron trajectories inside the lens.  The MCP/DLD will read out information of the arriving time and arriving location on the detector for each photoelectron, thus can be used to determine the electron energy and emission angle.  The bottom-right corner inset shows a zoom-in view of the MCP/DLD unit.}
\label{fig:ARTOF}
\end{figure}

\begin{figure}
\centering\includegraphics[width=1\textwidth]{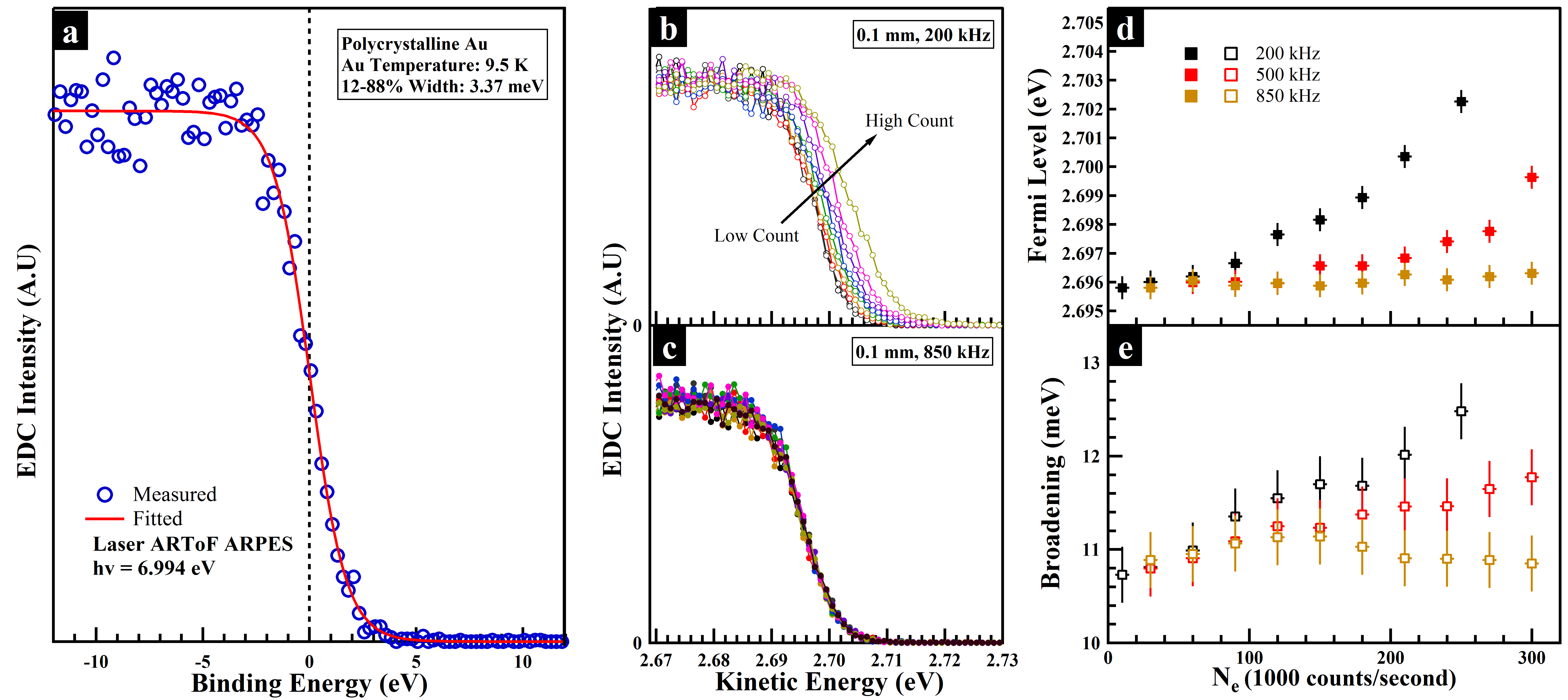}
\caption{ Testing results of the VUV-laser-based ARToF-ARPES system. (a).  EDC from measuring on a polycrystalline Au by using  6.994 eV laser light at 9.5 K.  The measured Fermi edge width is  $\sim$3.37 meV. After removing the contribution from temperature,  the overall instrumental energy resolution obtained is better than 1 meV.  (b) and (c) are the polycrystalline Au  EDCs taken at two different laser repetition rates. The fitted results of the Fermi level position and corresponding Fermi edge width are summarized in (d) and (e). It is clear that as one increases the operating repetition rate of the laser light,  the space charge effect can be strongly suppressed.}
\label{fig:Eresolution}
\end{figure}

\begin{figure}
\centering\includegraphics[width=1\columnwidth]{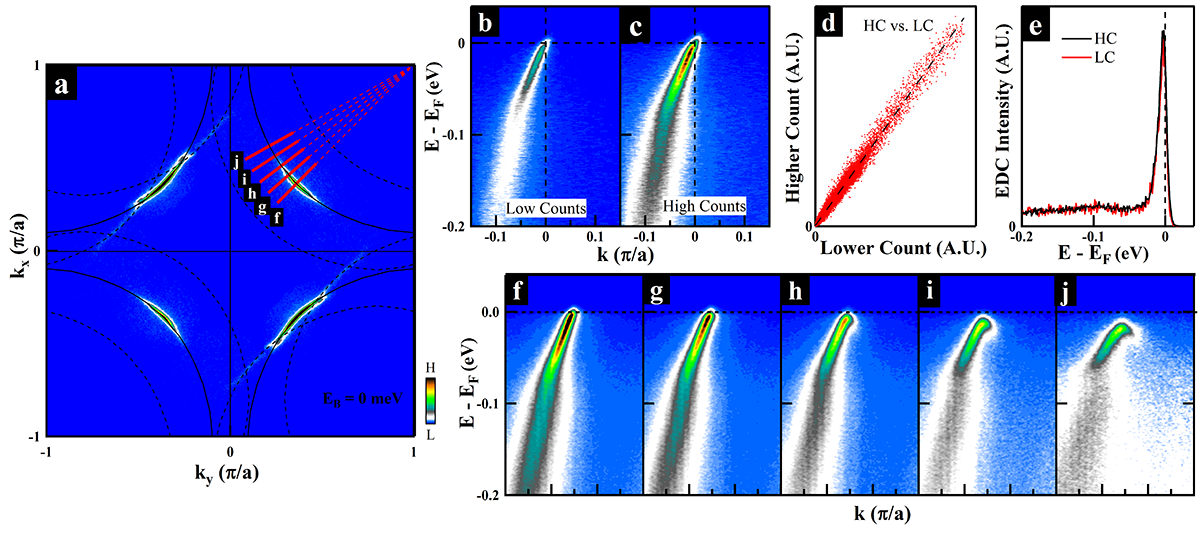}
\caption{Fermi surface and band structure of Bi2212 measured by VUV laser-based ARToF-ARPES.   (a).  Fermi surface measured by ARToF-ARPES with 6.994 eV laser light. The solid and dashed thin lines are  simulated Fermi surface and related superstructure replicas, respectively,  within the first Brillouin zone.    (b, c) are two band structures measured along the nodal direction with low count rate and high count, respectively.  (d) is the scatter plot (red scatters) showing the linear scaling relationship between the count rate at each pixel from both measurements. (e). EDCs  from (a-c) at the same Fermi momentum measured using low and high count rates. They match each other very well after scaling their intensity.   Both (d) and (e) indicate little  nonlinearity for the  ARToF analyzer.   (f-g).  Band structure taken along the momentum cuts shown in (a) that all extend to the ($\pi$, $\pi$) point. }
\label{fig:FigNonLE}
\end{figure}

%%Time-resolved
\begin{figure}
\centering\includegraphics[width=1\columnwidth]{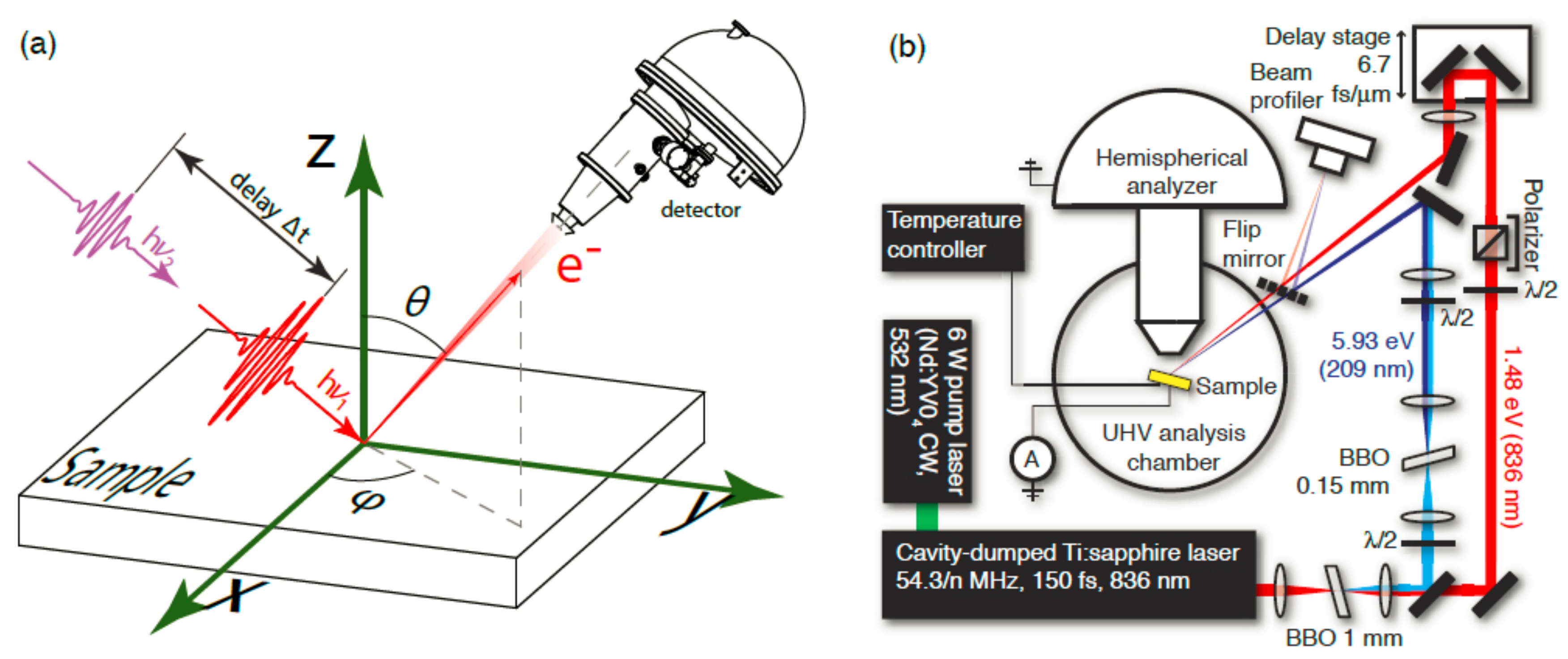}
\caption{(a) Schematic of a time-resolved ARPES. (b) A typical setup of time-resolved ARPES based on solid laser source\cite{Smallwood2012b}.}
\label{Fig7}
\end{figure}

%Scientific Applications
%Application: Eliashberg Functions
\begin{figure}
\centering\includegraphics[width=1\columnwidth]{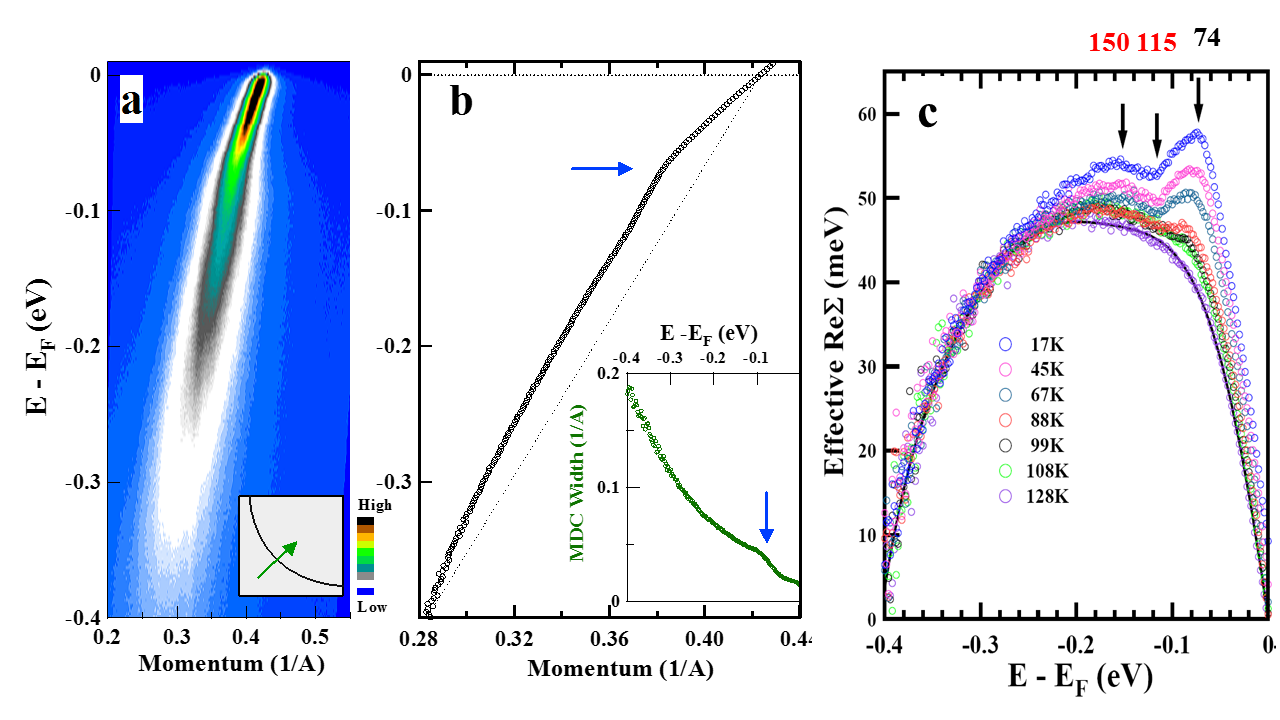}
\caption{Electronic structure of Bi2212 from VUV-laser ARPES\cite{WTZhang2008a}.  (a). Band structure of Bi2212 taken along the nodal direction.   (b). Quantitative dispersion and scattering rate obtained by MDC fitting of the original data (a).   (c).  The effective real part of electron self-energy at different temperatures.  It is clear that there are two high energy features at 115 meV and 150 meV in addition to the well-known 70 meV mode-like feature. }
\label{fig:NewModeinBi2212}
\end{figure}

\clearpage

\begin{figure}
\centering\includegraphics[width=1\columnwidth]{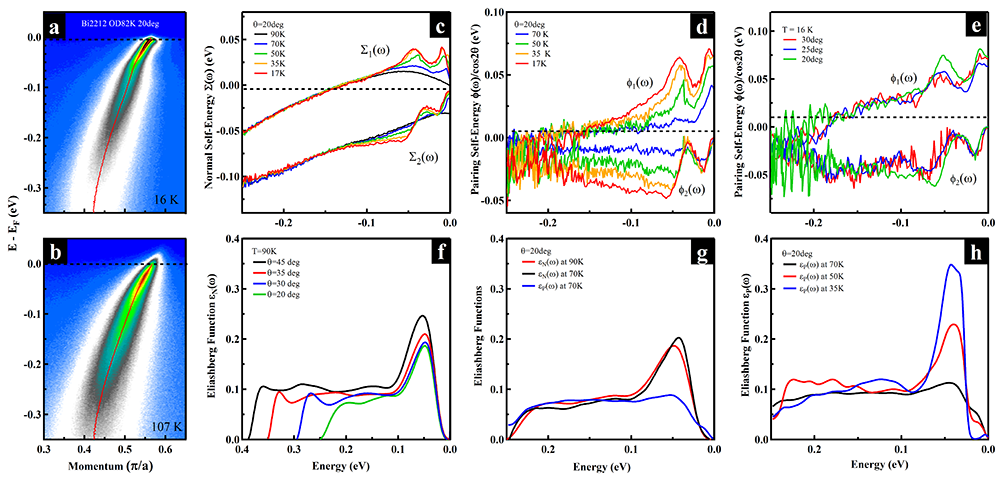}
\caption{Extraction of Eliashberg functions from VUV laser-based ARPES on Bi2212\cite{Bok2010}.   (a) and (b) are high resolution ARPES data  from over-doped Bi2212 with a T$_c$=82 K measured at 16 K and 107 K, respectively.   (c). Normal electron self-energy;  (d) is pairing self-energy at different temperatures for a given cut. (e) is pairing self-energy at different momentum cuts for a given temperature.   (f).   The extracted normal Eliashberg function  for different momentum cuts. (g). Extracted normal Eliashberg function and pairing Eliashberg function for Bi2212. (h). Extracted pairing Eliashberg function for Bi2212.
}
\label{fig:Eliashberg}
\end{figure}

%Application:Spin
\begin{figure}
\begin{center}
\includegraphics[width=1.0\columnwidth]{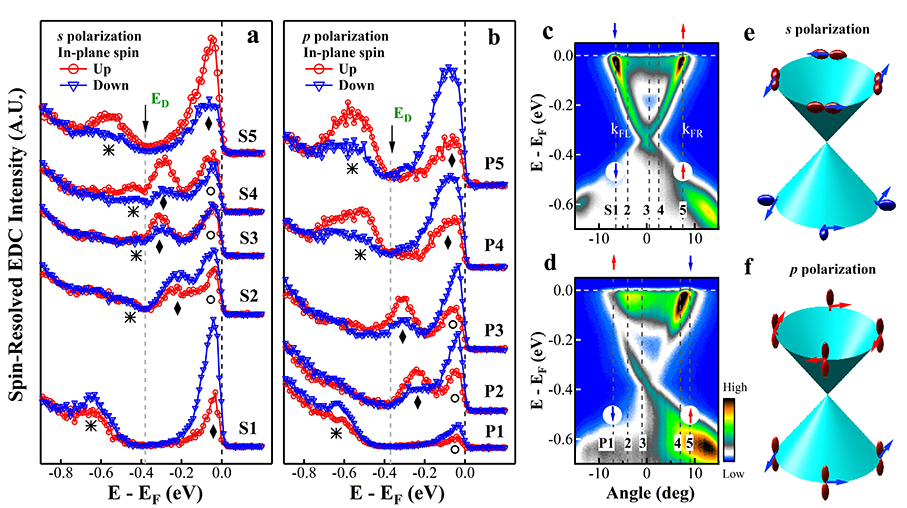}
\end{center}
\caption{ SARPES measurements of Bi$_2$Se$_3$ topological insulator under different polarization geometries\cite{ZJXie2014a}.  {a}. Spin-resolved EDCs at five representative momenta along the $\bar{\Gamma}\bar{K}$ momentum cut under the {\it s} polarization geometry. The corresponding momentum points are marked as the dashed lines in the band image measured from regular ARPES under the {\it s} polarization geometry  (c). The EDC  peaks corresponding to the bulk band, the upper Dirac cone and the lower Dirac cone are marked by empty circle, solid diamond and asterisk, respectively.    (b).  Spin-resolved EDCs at five representative momenta along the $\bar{\Gamma}\bar{K}$ momentum cut under the {\it p} polarization geometry. The corresponding momentum points are marked as the dashed lines in the band image measured from regular ARPES under the {\it p} polarization geometry.  (d). On the right side of \textbf{c}, the measured spin and orbital texture under the {\it s} polarization geometry is sketched for both the upper Dirac cone and the lower Dirac cone. On the right side of \textbf{d}, the spin and orbital textures under the {\it p} polarization geometry is sketched.}
\label{fig:SpinOrbital}
\end{figure}

%WuTe2 and ZrTe5

\begin{figure}
\begin{center}
\includegraphics[width=1.0\columnwidth,angle=0]{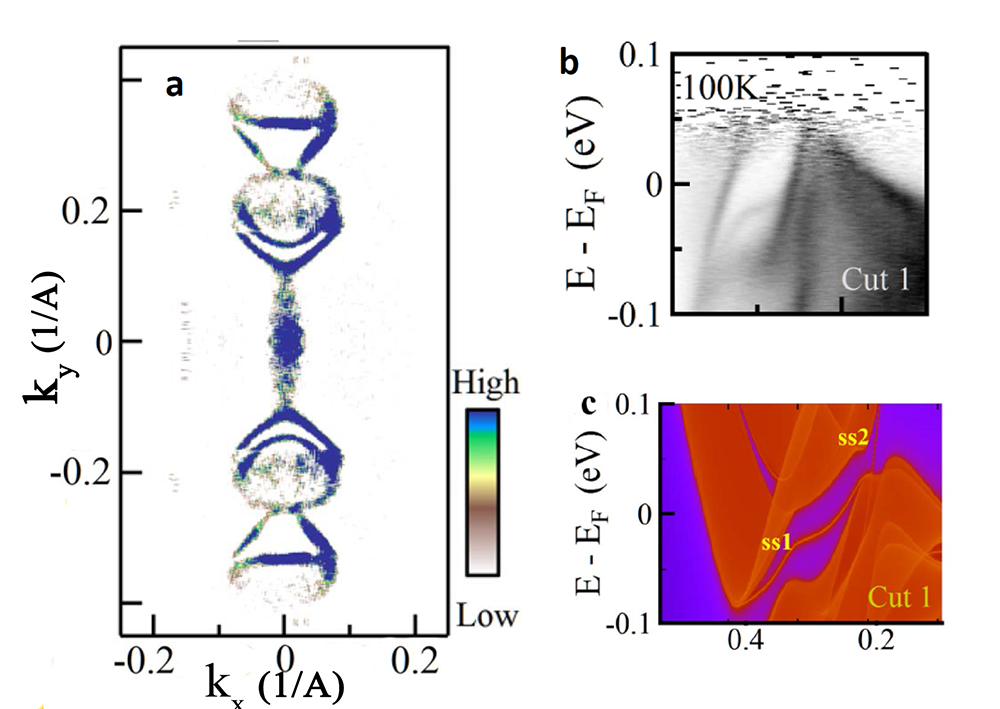}
\end{center}
\caption{Complete electronic structure of WTe$_2$ from VUV laser-based ARToF-ARPES measurements\cite{CLWang2016}.   (a). The measured complete Fermi surface of WTe$_2$.  (b).  Observation of the surface state band at 100 K. (c). Calculated band structure. }
\label{FigWTe2}
\end{figure}

\clearpage

\begin{figure}
\begin{center}
\includegraphics[width=1.0\columnwidth,angle=0]{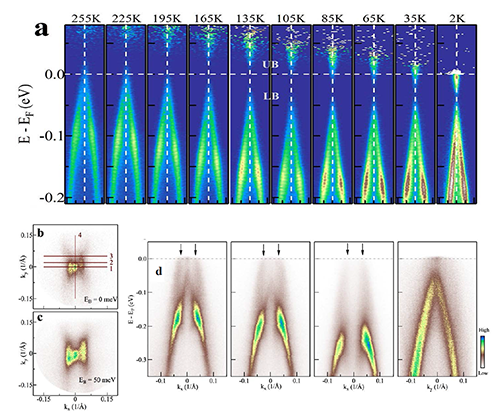}
\end{center}
\caption{Temperature-induced Lifshitz transition and topological nature of ZrTe$_5$ from VUV laser-based ARToF-ARPES measurements\cite{YZhang2017}.  (a). Temperature evolution of the band structure in ZrTe$_5$;   (b). Observation of weak topological insulator feature, the one-dimensional-like streaks, in some ZrTe$_5$ samples.}
\label{FigZrTe5}
\end{figure}

%Time-Resolved ARPES
\begin{figure}
\centering\includegraphics[width=1\columnwidth]{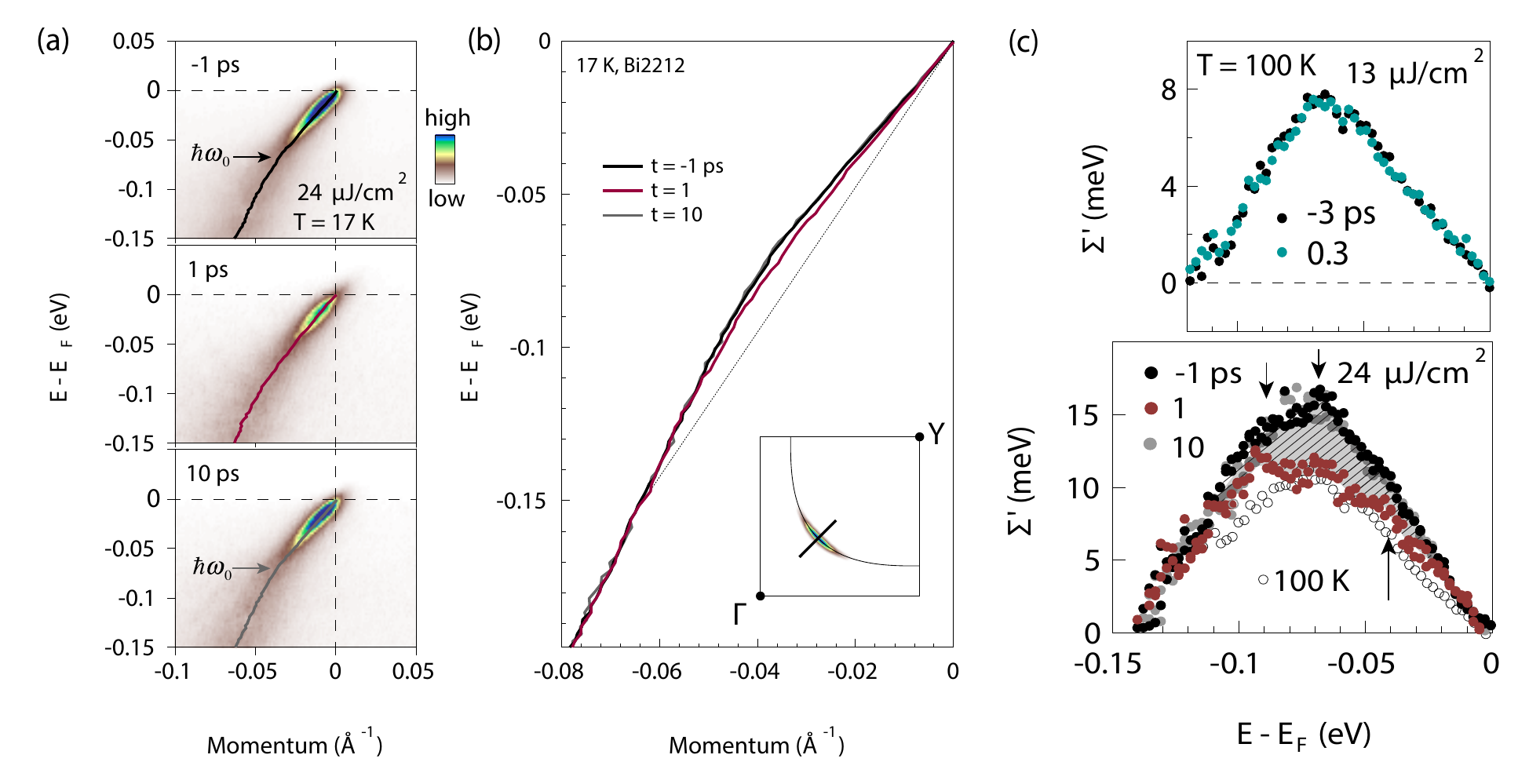}
\caption{Time-resolved ARPES data of an optimally-doped Bi2212 sample ($T_c$ = 91 K) measured along a nodal cut ($\Gamma$(0, 0) -- Y($\pi$, $\pi$) direction)\cite{WTZhang2014}.
(a).  Equilibrium (before pumping, $t=-1$ ps), transient (after pumping, $t = 1$ ps and $t = 10$ ps) photoelectron intensity (represented by false color) as a function of energy and momentum, for a pump fluence of 24 $\mu$J/cm$^2$. The bold solid black lines are the momentum distribution curve  dispersions at the corresponding delay time. The arrows mark the position of the kink at $\sim$70 meV.  (b).  MDC dispersions for different delay times ($-1$, $1$, and $10$ ps).  (c).  The effective real parts of the electron self-energy at 100 K (upper panel) and at 16 K (lower panel). }
\label{Fig13}
\end{figure}

\clearpage

\begin{figure}
\centering\includegraphics[width=1\columnwidth]{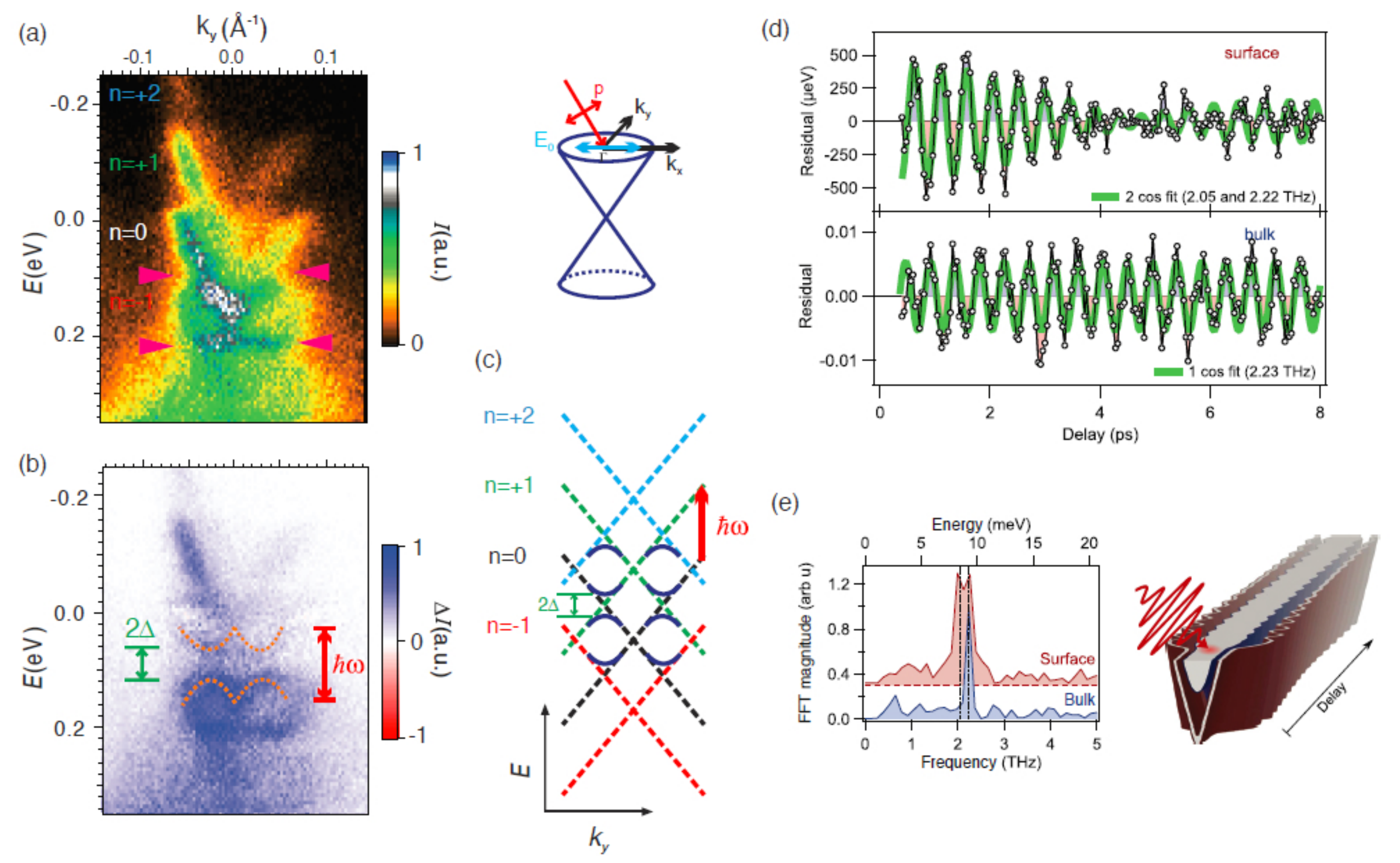}
\caption{Time-resolved ARPES data of a Bi$_2$Se$_3$ sample.  (a).  The time-resolved ARPES data through $\Gamma$ along the k$_y$ direction pumped with linear polarized pulse\cite{YHWang2013}.  (b).  The same data after subtracting the data at $t=-500$ fs\cite{YHWang2013}.  (c). Sketch of the side bands of different order\cite{YHWang2013}. (d).  Pump-induced electronic dynamics in surface bands (upper panel) and bulk bands (lower panel)\cite{Sobota2014}.  (e).  Fourier transforms of the spectra shown in (d) (left) and cartoon depiction of the optical excitation of coherent phonons in the bulk and surface bands (right)\cite{Sobota2014}. }
\label{Fig14}
\end{figure}

%%End Figrues

\end{document}